\documentclass{ametsocV6.1}

\title{On the limits of Taylor’s hypothesis in forest clearcut flow: Effects of random sweeping and an elliptic model analysis}

\authors{Subharthi Chowdhuri,\aff{a}\correspondingauthor{Subharthi Chowdhuri, subharthi.chowdhuri@luke.fi}
Ivan Mammarella,\aff{b} 
Olli Peltola,\aff{a} 
}

\affiliation{\aff{a}{Natural Resources Institute Finland (Luke), Latokartanonkaari 9, Helsinki, 00790, Finland}\\
\aff{b}{Institute for Atmosphere and Earth System Research (INAR)/Physics, Faculty of Science, University of Helsinki}\\
}

\abstract{Taylor's hypothesis (TH) converts temporal observations to spatial information of the flow while carrying out measurements on a micrometeorological tower. Other than TH, there exists a more general elliptic model, which converts time to space by focusing on the geometry of the space-time correlation function. In elliptic model, TH is recovered when the space-time correlation functions are straight lines and when TH is invalid, they are approximated as elliptic curves. To test whether TH or an elliptic model was appropriate for a highly heterogeneous forest clearcut flow, we examined the geometry of the space-time correlation function of temperature by using an extensive distributed temperature sensing (DTS) dataset collected during daytime convective conditions at a height of 3.1 m above the clearing. This was complemented with an eddy covariance (EC) dataset that measured the turbulence characteristics. When the mean wind was parallel to a nearby forest edge, the DTS-derived space-time correlation function of temperature fluctuations resembled elliptic curves, rather than straight lines as predicted by TH. Due to large turbulence intensities, the curvatures in the space-time correlation contours were caused by the random sweeping events associated with large-scale eddies that invalidated the frozen turbulence assumption in TH. The velocity scale of these sweeping events correlated with the turbulence kinetic energy of the clearcut flow, thereby lending support to the random sweeping hypothesis. Upon converting time to space through an elliptic and TH-based scaling, our results demonstrated that both temporal DTS and EC temperature measurements were impacted by the random sweeping events.} 

\begin{document}

\maketitle

%
%
%
\statement
The near-surface turbulence measurements serve as a cornerstone in land-atmosphere interaction research. Sans spatial information, to associate these temporal measurements with spatial scales, one applies Taylor’s hypothesis (TH) by assuming that the space-time conversion can be achieved through the local mean wind speed. However, this hypothesis fails when the turbulence intensities are large and random sweeping effects associated with large scale eddies invalidate the frozen turbulence assumption in TH. To account for this sweeping effect, we employ an elliptic model of space-time correlation in a forest clearcut flow. The results suggest that in a highly turbulent and heterogeneous forest clearcut flow, the relationship between space and time is more closely approximated as an ellipse rather than a simple linear one.  
%
%

%
\section{Introduction}
\label{Intro}
Across the globe, the micrometeorological towers routinely monitor the vertical turbulent fluxes of  heat, matter (e.g. moisture, $\rm CO_{2}$) and momentum under the aegis of various programs such as FLUXNET, AmeriFlux, NEON, and ICOS \citep{baldocchi2001fluxnet,novick2018ameriflux,metzger2019neon,heiskanen_integrated_2022}. The data collected from these efforts serve as a cornerstone to understand land-atmosphere interaction, energy budget of the Earth's surface, and carbon and water exchanges between the ecosystem and atmosphere. Although the turbulent eddies facilitating the vertical transport exist physically in space, spatial information of atmospheric turbulent flows are rarely available and thus the instruments mounted on micrometeorological towers register the turbulent variables over time. To circumvent this issue, Taylor's frozen turbulence hypothesis (TH) is used to convert the temporal information to spatial ones. This hypothesis states that the turbulent structures move past the measurement location with a convective velocity in a frozen state, i.e. their characteristics change slowly compared to the speed at which the mean flow transports them along \citep{taylor1938spectrum}. 

The validity of TH in atmospheric flows has been studied previously by utilizing theoretical \citep{wyngaard1977taylor,wilczek2014note}, numerical \citep{horst2004hats,higgins2012effect}, and novel experimental techniques \citep{cheng2017failure,han2019applicability,hilland2024systematic}. Notwithstanding this progress, a methodological challenge persists in how the convective speeds of the turbulence structures should be evaluated, which is central to the assessment of TH. Some studies have utilized the peak time scales associated with the temporal cross-correlation functions computed between different locations over space \citep{shaw1995wind,han2019applicability}. Under the assumption that the structures take longer time to reach distant spatial locations, \citet{krogstad1998convection} showed that a linear relationship could be found between the peak time scales and spatial locations, whose slope provided an estimate of the bulk convective speeds. This method has been utilized on roughness sublayer and atmospheric surface layer flows \citep{powell1974investigation,shaw1995wind,horst2004hats,han2019applicability} and the results often yield convective speeds larger than the mean wind speed. On the other hand, \citet{su2000two} utilized the large eddy simulation data of a dense canopy flow and they found that the temporal and spatial peaks of cross-correlation functions varied differently with space and time. This effectively produced two convective velocities where the one computed over the temporal domain appeared larger than the one computed over the spatial domain. Later, \citet{cheng2017failure} utilized the methodology of \citet{del2009estimation} to compute the convective speeds of temperature structures by using the Fourier phases of the spatial signal obtained from distributed temperature sensing measurements and large eddy simulations. They found that at larger scales this convective speed was nearly equal to the mean wind speed and decreased as a power of $-1/3$ with increasing wavenumbers. This conclusion was qualitatively similar to \citet{higgins2012effect}. By utilizing both Raman lidar and large eddy simulation data and employing a threshold on the temporal cross-correlation functions of humidity, \citet{higgins2012effect} found that at larger scales the structures nearly moved with the mean wind speed.  

From this brief review it is clear that different methodologies yield different outcomes, which forces one to ask how the validity of TH should be assessed in atmospheric flows. For instance, if the convective speeds of turbulence structures are larger or smaller than the mean wind speed, is it sufficient to claim that TH is invalid? \citet{cheng2017failure} argued that this is not a strong test as it is often the frozen turbulence assumption in TH that is not valid. In fact, from a turbulent jet experiment, \citet{wills1964convection} argued that when the frozen turbulence assumption does not hold, the morphology of space-time correlation contours are not straight lines as assumed in TH, rather they appear as closed-form curves. Under such conditions, \citet{wills1964convection} showed that there will be differences between the convective speeds estimated from the spatial and temporal domains. This was precisely the case with \citet{su2000two}. \citet{su2000two} mentioned that, without showing any results, the space-time correlation contours of velocity and scalar fluctuations from their large eddy simulation data of canopy flows were of elliptic shapes. After noticing these issues, \citet{everard2021sweeping} advocated the use of an elliptic model to assess the validity of TH in atmospheric flows. 

The elliptic framework was introduced by \citet{he2006elliptic} and \citet{zhao2009space} to model the morphology of space-time correlation contours in turbulent flows. In the case of a statistically homogeneous and stationary flow, the space-time correlation of any turbulent variable is a function of $r$ and $\tau$, where $r$ is the spatial separation and $\tau$ is the temporal lag. By expanding the space-time correlation function around the origin as a Taylor series expansion and only retaining up to the second order terms, \citet{zhao2009space} showed that an elliptic relationship can be obtained to express the form of space-time correlations as,
\begin{equation}
    (r-U_e\tau)^2+V^2\tau^2=\lambda_x^2\left[1-R_{xx}(r,\tau)\right],
    \label{em2}
\end{equation}
where $\lambda_x$ is the Taylor microscale and $R_{xx}(r,\tau)$ is the space-time correlation function for a turbulent variable $x$. In a given turbulent flow ($\lambda_x$ is fixed) and for a specific contour level of $R_{xx}(r,\tau)$, the right-hand-side of Eq. \ref{em2} is a constant. Moreover, in Eq. \ref{em2}, $U_e$ is the convective speed of the turbulence structures and $V$ is the sweeping speed. The sweeping speed is associated with Kraichnan's random sweeping hypothesis \citep{kraichnan1964kolmogorov}, which dictates that the large scale turbulent eddies move the smaller scale eddies randomly across space without significant distortion. The presence of sweeping effects essentially invalidate the frozen turbulence assumption in TH as the structures no longer move as a frozen pattern along the streamline. In other words, under such conditions, TH does not remain as a good approximation of the space-time correlation structure. 

\begin{figure*}[h]
\centering
\hspace{-5cm}
\includegraphics[width=1\textwidth]{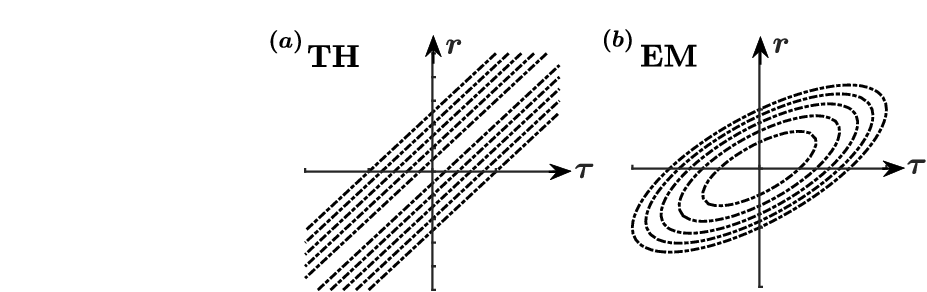}
  \caption{A schematic is used to illustrate the differences between the space-time correlations (shown as dash-dotted lines) obtained from (a) Taylor's hypothesis (TH) and (b) elliptic model (EM). Here, $r$ denotes the spatial separations and $\tau$ are the temporal lags. TH predicts the iso-correlation contours to be straight lines while EM predicts them to be ellipses. The equations representing these family of curves are $r-U_e\tau=C$ and $(r-U_e\tau)^2+V^2\tau^2=C$, respectively, with $C$ being a constant.}
\label{fig:1}
\end{figure*}

In this elliptic framework, one recovers TH when the sweeping speeds become negligible as compared to $U_e$. At an other extreme, when $U_e$ becomes very small, one obtains the classic case of a zero mean flow where sweeping effects solely dominate \citep{tennekes1975eulerian}. By nature, the elliptic model is supposed to hold for small values of $r$ and $\tau$. However, by assuming a scale-invariance exists, the elliptic model could be extended to larger scales \citep{zhao2009space}. To illustrate the difference between TH and elliptic model, we show a schematic in Fig. \ref{fig:1}. This schematic represents how the space-time contours look like when compared between TH and the elliptic model. In TH, the space-time correlation contours are a family of straight lines, while in the elliptic model the same appear as ellipses. From this geometrical point of view, TH is a first-order approximation of $R_{xx}(r,\tau)$ while the elliptic model introduces second-order terms to account for any deviations from the straight line \citep{he2006elliptic,zhao2009space}. 

Ideally, to test the elliptic model, one should have a spatio-temporal data that cover an extensive range of scales in both spatial and temporal domain. In reality, it is difficult to obtain such dataset from both experiments and numerical simulations due to measurement limitations and challenges associated with storing a large volume of data. Given this practical constraint, the success of elliptic model is assessed by investigating how well Eq. \ref{em2} collapses the temporal cross-correlation curves of turbulent variables at few spatial locations by utilizing high resolution temporal data. This remains true for various data sources on which the elliptic model has been applied. For instance, direct numerical simulation data of turbulent shear flows \citep{zhao2009space}, or particle image velocimetry and thermistor data from Rayleigh-B{\'e}nard convection experiments \citep{he2010small,zhou2011experimental}. Only for turbulent shear flows, \citet{zhao2009space} showed that, for the velocity components, one could theoretically estimate $U_e$ and $V$ from the governing Navier-Stokes equations. For other scenarios, such as when temperature is considered, two methods exist to compute them empirically. In the first method, two linear relationships are utilized between the peak positions of temporal (spatial) cross-correlation curves and spatial (temporal) separations to estimate $U_e$ and $V$ \citep{he2010small,zhou2011experimental,wang2014trpiv}. In the second method, the information in the temporal domain is utilized to compute the same \citep{hogg2013reynolds,he2015reynolds,musilova2017reynolds}. Previous studies have used these two methods interchangeably and therefore no systematic comparison exists between the two.

\citet{everard2021sweeping} applied this elliptic model on the thermocouple temperature data collected at a few spatial points in a vineyard canopy flow. They empirically determined its parameters by utilizing the first method and showed that the $U_e$ values of temperature structures were lower than the mean wind speed. This finding contradicted with the previous studies that utilized the peaks of the temporal cross-correlation curves \citep{shaw1995wind}. \citet{everard2021sweeping} claimed that this difference existed because of random sweeping effects, which decorrelated the temporal measurements faster and hence their peak positions appeared biased. \citet{everard2021sweeping} also found that the sweeping speeds scaled with the turbulence kinetic energy of the flow. This scaling relationship revealed that the sweeping effects were indeed associated with energetic large scale eddies of the flow, thereby lending support to Kraichnan's random sweeping hypothesis. Furthermore, \citet{everard2021sweeping} argued that, in the presence of random sweepings, the smaller scale eddies decorrelated faster in the temporal domain than what one would obtain in the corresponding spatial domain. As per \citet{everard2021sweeping}, to accommodate this disparity between the two domains, caused by random sweepings, the convective speeds of smaller scale eddies could appear as scale-dependent, which was consistent with the findings of \citet{cheng2017failure}. Despite these insights, given the limited spatial resolution of their dataset, \citet{everard2021sweeping} did not explicitly present the geometrical structure of the space-time correlation function and presumed that the elliptic shape holds. Therefore, further verification was needed to ascertain if the space-time correlation structure of temperature field in atmospheric flows could indeed be modeled as elliptic curves. 

Later on, \citet{han2022applicability,han2022predictive} conducted a study over a homogeneous grassland on the applicability of elliptic model by using the distributed temperature sensing measurements of \citet{cheng2017failure}. These studies did not focus on the scaling relationship between the sweeping speeds and the turbulence kinetic energy of the flow, rather they concentrated on the methodological aspects and geometrical structure of the space-time correlation function of temperature. For instance, \citet{han2022applicability} found that differences existed in what method one applies to empirically obtain the elliptic model parameters in atmospheric flows, and they recommended to use an averaged value of $U_e$ and $V$ obtained from the two methods. They could not provide any physical reason to do this, as they mentioned, \emph{the differences between the results obtained by the two methods are not obvious} (P. 076602-9). On the other hand, for the same dataset of \citet{cheng2017failure}, \citet{han2022predictive} found that the geometry of the space-time correlation function of temperature deviated from the standard form of ellipse. They attributed this deviation to the complexity and uncontrollable conditions in atmospheric flows, while also acknowledging that they only utilized a few 30-min periods and therefore their results were not statistically representative.

To date, in the context of atmospheric flows, these are the only two studies that have employed the elliptic model as an alternative to TH for studying the relationship between space and time. Both of these studies focus on homogeneous surface conditions with varying degrees of roughness (grassland vs. vineyard canopy), while, in reality, it is more common to encounter heterogeneous surface conditions \citep{bou2020persistent}. As a result of this scarcity, several questions still remain unresolved. First, it is not yet clearly demonstrated whether in an atmospheric flow, one could model the space-time correlation function as ellipses. It is prudent to ask if deviations from elliptic shapes exist and whether those are more significant in heterogeneous conditions as opposed to homogeneous flows. Second, in a strict sense, $V$ is a fitting parameter in the elliptic model and since it is related to the aspect ratios of the space-time correlation contours, $V$ cannot be a negative quantity. Only when it is established that $V$ scales with the turbulence kinetic energy (TKE) of the flow (serves as a velocity scale of large-scale eddies), one can assign physical meaning to this quantity. Under such scenario, $V$ can be interpreted as an effective velocity scale representing the typical magnitude of random sweeping motions induced by large-scale turbulent structures, contributing to temporal decorrelation. Except \citet{everard2021sweeping}, in the context of geophysical flows, we are not aware of any study where it has been established that $V$ and TKE are intimately connected to each other, let alone for a heterogeneous flow. Third, from a practical standpoint, it is important to address why there are differences in $U_e$ and $V$ when estimated from the two alternative methods. This is necessary because few studies in engineering flows and only one study in atmospheric flow have used these two methods interchangeably with no systematic comparison and therefore, no explanation exists. 

To address these pertinent issues, we focus on a highly heterogeneous forest clearcut flow. This choice allows us to expand the scope of elliptic model to more realistic surface conditions, characterized by heterogeneity. Clearcutting is a widely utilized forest management practice to collect timber and in some cases also to restrict the spread of wildfires by limiting the access to the fuel \citep{ortega2024modeling}. In many respects (e.g. high turbulent intensities), forest clearcut flows share resemblance with roughness sublayer flows \citep{flesch_wind_1999}. Despite so, as compared to homogeneous roughness sublayer flows, it is not evident to what degree the random sweeping hypothesis would be valid and whether the sweeping speeds could be uniquely related to the turbulence kinetic energy. Accordingly, we define our research questions for this study as follows:
\begin{enumerate}
    \item Is TH a good approximation of the space-time correlation structure in a forest clearcut flow and if not, how important are the sweeping effects and could those be explained by the turbulence kinetic energy of the flow?
    \item Can the space-time correlation curves in a forest clearcut flow be approximated as ellipses?
    \item For practical purposes, which method one should use while computing the parameters of the elliptic model and what could be the reason if differences exist between the estimates?
    \item Do the routine eddy covariance measurements feel the impact of random sweeping events?
\end{enumerate}

To tackle these objectives, we employ an extensive dataset encompassing both distributed temperature sensing (DTS) and eddy covariance (EC) measurements, collected during a five-month field campaign in a forest clearcut. The DTS measurement technique uses fiber optic cables and relies on Raman scattering principle to convert the back-scattered energy from the emitted light pulses to temperature values at specific locations in space \citep{selker2006distributed,peltola2021suitability}. By laying out these cables horizontally, we obtain temperature data at a fine enough spatial (of the order of tens of cm) and temporal resolutions (of the order of seconds). We restrict our analysis to daytime convective periods when the temperature signals were sufficiently strong, in order to minimize the effect of noise on the results. The rest of this article is organized in three different sections. In Sect. \ref{mat and met} we describe our measurement set up and methodologies being adopted to verify the elliptic model. In Sect. \ref{results}, we present and discuss our results, and in Sect. \ref{conclusion} future research directions are outlined.

\section{Data and Methodology}
\label{mat and met}
\subsection{Site description}
\label{site}
\begin{figure*}[h]
\centering
\includegraphics[width=1\textwidth]{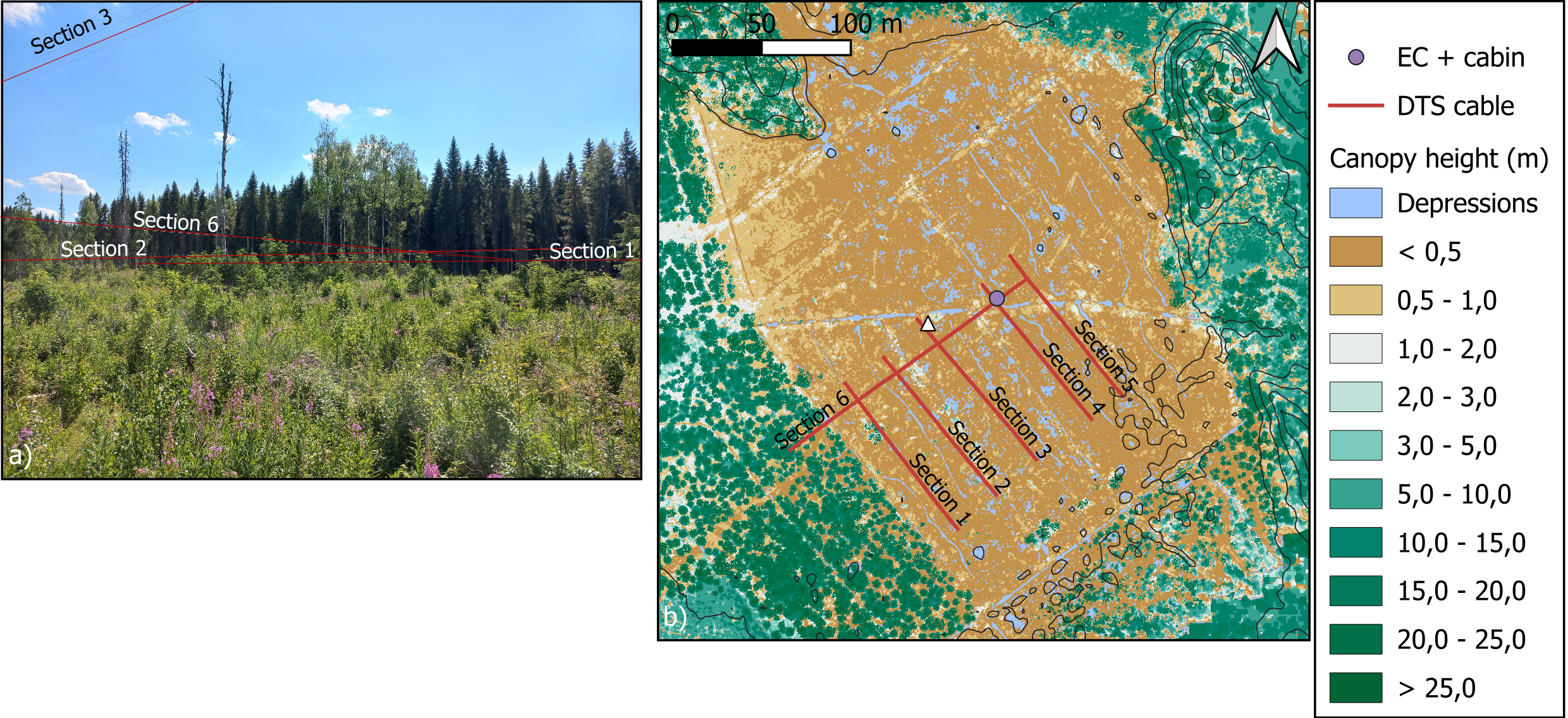}
  \caption{(a) Image and a (b) map of the study site. The image was taken towards South-West on June 28 2024 at the location of the white triangle in (b). Different DTS cable sections are highlighted with red lines. In (b) the black contours show the topography with 2-m increments based on the National Land Survey of Finland Topographic Database (data retrieved: 10/2023). The canopy height map was derived from a drone flight conducted during September 27 2022 and complemented with canopy height model of Finnish Forest Center (data retrieved: 12/2024). Note that the canopy height shown in the map does not fully reflect the conditions at the clearing during the study period due to the rapidly developing vegetation cover.}
\label{fig:map}
\end{figure*}

The measurement campaign took place between May 17 and September 10 2024 at Ränskälänkorpi study site (61$^{\circ}$11$^{\prime}$ N, 25$^{\circ}$16$^{\prime}$ E, 144 m above sea level), which is located on a peatland that was drained before 1960's for forestry purposes \citep{laurila_set-up_2021}. A Norway spruce (\textit{Picea abies}) stand growing on the site was clear-felled  during late winter - spring of 2021 creating a 6.4 ha clearcut area (see Fig. \ref{fig:map}) and Norway spruce seedlings were planted after the clearcutting. Few isolated groupings of living and dead trees were left untouched on the clearcut area and logging debris (e.g. branch piles) were left on site. In general, the terrain of the clearcut is flat, but its surface exhibits prominent small scale undulations due to e.g. drainage ditches and banks. After clearcutting, pioneer species, such as fireweed (\textit{Chamaenerion angustifolium}), downy birch (\textit{Betula pubescens}) and raspberry (\textit{Rubus idaeus}), have spread on the area and created a varying vegetation cover on the clearing (see Fig. \ref{fig:map}a). During the study period, the height of these plants varied approximately between 0.5 m and 1.5 m, depending on location and timing of the growing season. In summary, the clearcut area is a complex mosaic of isolated trees, pioneer species, and logging debris creating a rough surface for the flow.

The clearcut is surrounded by a forest dominated by Norway spruce (\textit{Picea abies}) with height between 20 m and 25 m depending on the location (see Fig. \ref{fig:map}b). Leaf area index of the forest in SW direction of the clearing is approximately 2.4 m$^{2}$ m$^{-2}$. The forest edges create abrupt changes in canopy height and hence in surface roughness.

\subsection{Measurement setup}
\label{measurements}
The spatial and temporal variability of air temperature above the clearcut was measured with the distributed temperature sensing (DTS) technique \citep{thomas_optical_2021}. The DTS instrument (Ultima-S, 5 km variant, Silixa Ltd., Hertfordshire, UK) was housed in a temperature controlled cabin in the middle of the forest clearcut (see Fig. \ref{fig:map}b). The measurements were conducted using double-ended configuration and signals from both ends of the fiber-optic cable (forward and reverse signals) were stored for later processing (see Sect. 2\ref{DTS_processing}). The DTS instrument provided observations with 1.2 s temporal and 12.7 cm spatial resolution along the attached cable. Note that in double-ended configuration both ends are measured sequentially and hence after processing the temperature observations were available with 2.3 s resolution.

Two calibration locations were created next to the DTS instrument and both ends of the cable were guided through them providing altogether four calibration sections in the data. The calibration sections were at minimum 10 meters long. Cold calibration point was created by digging the cable to 0.7 m depth in the peat soil. This depth was below the soil water level and hence the temperatures at that depth did not vary rapidly. Typically, the temperatures were below the air temperature and varied between 2.9 $^\circ C$ and 10.7 $^\circ C$ during the campaign. Warm calibration point was created with 50-l calibration bath filled with water and it was located in the temperature controlled cabin. Water heater was installed on June 28 and it was set to keep the water at 30 $^\circ$C. Prior to that the water temperature followed indoor air temperature in the temperature controlled cabin. Temperatures at the calibration locations were monitored with reference thermometers (PT-100) supplied with the DTS instrument.

A white, thin (outer diameter 0.9 mm) aramid reinforced 50 $\mu$m multimode fiber-optic cable (AFL Telecommunications, Duncan, SC 29334, US) was utilized in this study. The cable was arranged in six sections and along each section the temperature values were doubly measured. This was because the measurements were made with a single continuous cable traversing back and forth the different sections (Fig. \ref{fig:map}). Sections 1 to 5 were aligned with the forest edge located in South-West direction of the clearcut and the section lengths were between 97 m and 103 m. The cable sections 1, 2, 3, 4 and 5 were approximately at a 20 m, 47 m, 75 m, 115 m and 137 m distance from the forest edge in SW direction, respectively. The sixth section was perpendicular to the other cable sections and to the forest edge and its length was 164 m, out of which 27 m was within the forest. Therefore, the total cable length was 1478 m.

Due to heterogeneity in the surface conditions (see Sect. 2\ref{site}), the cable height above the soil changed from one location to another. To avoid any terrain-induced curvature effects on the cable orientation, we decided not to follow the local terrain but to fasten the cable to wooden support structures at a constant height above sea level. This height was identified at each location with a handheld GNSS device (Geo 7X, Trimble Inc., USA), utilized to position the support structures in the area. At the location of the eddy covariance (EC) station, the cable height was around 3.1 m above the soil (see Fig. \ref{fig:map}b). However, the local DTS measurement height varied and was approximately between 2 to 4 m above the soil and between 0.5 to 2.5 m above the local roughness elements (e.g. vegetation). Moreover, occasionally the cable was also surrounded by the few isolated trees left at the clearcut. For practical purposes, we ignore these local height variations and consider the height of the DTS cable to be 3.1 m above the soil in order to be consistent with the EC measurements. 

To quantify the turbulence conditions over such a complex surface, the EC equipment consisted of a 3-D sonic anemometer (uSonic-3 Cage MP, METEK GmbH, Germany) and an Infrared gas analyzer (LI-7200RS, LI-COR Biosciences, NE, USA). These together measured the three wind velocity components, sonic temperature, and carbon dioxide and water vapor mixing ratios at the height of 3.1 m above the soil. All the EC data were logged with 10-Hz frequency and for our purposes, only the 3-D sonic anemometer data sufficed. See more details on the EC measurement setup and flux footprint calculations in \citet{tikkasalo2025eddy}.

\subsection{DTS data processing}
\label{DTS_processing}
DTS data were post-field calibrated with Python (3.11.9) scripts and the processing followed \citet{des_tombe_estimation_2020}, \citet{van_de_giesen_double-ended_2012} and \citet{peltola_probing_2022} with slight differences. Temperature can be estimated from the observed Stokes ($P_s$) and anti-Stokes ($P_{as}$) signals with the following equation \citep{van_de_giesen_double-ended_2012}:
\begin{equation}
    T_x(s,t) = \frac{\gamma}{\ln{\frac{P_s(s,t)}{P_{as}(s,t)}}+C(t)+\int_0^s\alpha(s')ds'},
    \label{eq:DTS_calibration}
\end{equation}
where $T_x=T_x(s,t)$ is the calibrated temperature at location $s$ along the fiber-optic cable at time step $t$, $\gamma$ and $C(t)$ are calibration parameters, and $\int_0^s\alpha(s')ds'$ describes the differential attenuation of $P_s$ and $P_{as}$ along the fiber. This integral was estimated from the forward and reverse signals for each location $s$ separately, corresponding to individual 30-min periods \citep{van_de_giesen_double-ended_2012}. Thereafter, at each time step, the calibration parameters $\gamma$ and $C$ were estimated for the forward and reverse signals by fitting Eq. \ref{eq:DTS_calibration} to $P_s$ and $P_{as}$ observed with the DTS instrument and $T$ measured with the reference sensors. Although, in theory, $\gamma$ should be constant, here it was treated as a fitting parameter varying in time. These parameters were used to calculate separate estimates of $T_x(s,t)$ from the forward ($T_f$) and reverse ($T_r$) signals. Since $T_f$ and $T_r$ were measured sequentially, $T_r$ data were linearly interpolated to match $T_f$ time stamps. 

Noise estimates ($\sigma_T$, 1-$\sigma$ variability related to random noise) of $T_f$ and $T_r$ observations were computed for each 30-min period as,
\begin{equation}
    \sigma_T^2(s) = \frac{\overline{\left(T_x(s,t)-T_x(s+\Delta s,t)\right)^2}}{2},
    \label{eq:DTS_noise}
\end{equation}
where $T_x$ represents either $T_f$ or $T_r$, $\overline{( )}$ denotes temporal averaging and $\Delta s$ is the spatial step along the cable (0.127 cm). In this equation, we assume that the variability between consecutive data points in space is dominated by noise. The calibrated temperature $T$ at each $t$ and $s$ was calculated as a weighted average of $T_f$ and $T_r$ where inverse of noise estimates were used as weights \citep{des_tombe_estimation_2020}. Then the noise estimates for $T$ were calculated following noise propagation of weighted averaging. The spatial resolution of the data were reduced to 30 cm with spatial aggregation using weighted averaging as above and noise estimates for these spatial aggregates were calculated using noise propagation as above.

After this calibration procedure, we obtained calibrated $T$ at each time step $t$ and location $s$ along the cable. The $T$ data measured along the cable were converted to physical coordinates (ETRS89/TM35FIN coordinate system) using the measured cable locations (Sect. \ref{measurements}). Since there were double cabling at each physical location, the double measurements made at each location were further averaged using weighted averaging as above with the noise estimates obtained from the procedure described above. By utilizing this procedure, we obtained calibrated $T$ data with 2.3 s temporal and 30 cm spatial resolution throughout the measurement campaign at well-defined physical ($x$,$y$,$z$) locations (red lines in Fig. \ref{fig:map}) over the study area. These calibrated temperature values were compared against the sonic anemometer measurements at the clearcut showing satisfactory agreement between the two (not shown).

To use the DTS and EC datasets, we focus on buoyancy-dominated conditions when the temperature signals were sufficiently strong (see Fig. \ref{fig:A1} of Appendix A for further discussion). In particular, we selected those 30-min periods when there was no rain, incoming shortwave radiation was larger than 300 $\rm W \ m^{-2}$ and it was not strongly varying within the 30-min periods, the sensible heat flux ($>50 \rm \ W m^{-2}$) and variance of temperature ($> 0.1 \rm  \ K^2$) were sufficiently large, air temperature did not exhibit strong trends, and the direction of the mean wind was within $\pm 20^{\circ}$ of the DTS cable positions (see Fig. \ref{fig:map}b). 

For the present study, we restrict ourselves to the periods when the wind direction was parallel to the forest edge, i.e. when it coincided with the DTS cable sections 1 to 5. We exclude the DTS temperature data from section 6 for this study. The section 6 was aligned perpendicular to the forest edge (see Fig. \ref{fig:map}b). Due to direct influences of the upstream forest canopy on the clearcut flow, it introduced complications while applying the elliptic method for the perpendicular-wind periods. This particular flow condition deserves to be studied separately and will be reported in a future work. 

After applying all these conditions, we ended up with a total of nearly 56 hours of DTS and EC data, or in other words, 113 half-hourly periods. The DTS dataset was used to estimate the space-time correlation curves by using the temperature signal as a tracer, while the EC dataset provided information on the turbulence intensities, kinetic energies, and sonic temperatures. Corresponding to the EC data, $u$, $v$, and $w$ indicate the streamwise, cross-stream, and vertical velocities, while $U$ indicate the mean wind speed. After employing a double-coordinate rotation, the turbulent fluctuations in these quantities were computed through linear detrending. 

\subsection{Estimation of the elliptic model parameters}
\label{speed}
It is important to recognize that the elliptic model is strictly valid for a statistically homogeneous and stationary flow \citep{he2015reynolds}, while here we deal with highly heterogeneous surface conditions. In a turbulent Rayleigh-B{\'e}nard convection experiment, to check the statistical homogeneity assumption in the elliptic model for temperature, \citet{he2015reynolds} evaluated the temporal probability density functions (PDFs) and auto-correlation functions of temperature fluctuations at various spatial points. By showing that these PDFs and auto-correlation curves did not display any dependence on the spatial locations, \citet{he2015reynolds} concluded the flow to be statistically homogeneous. 

We do the same exercise for all the five DTS cable sections, where we plot the temporal temperature fluctuation PDFs and auto-correlation curves at several locations along the cable (see Figs. S1 and S2 in the Supplementary Material). The PDFs and auto-correlation curves collapsed reasonably well among different locations despite having highly heterogeneous surface conditions. For all the locations along the cable, the temporal temperature PDFs were asymmetric and skewed towards the positive fluctuations, a characteristic commonly associated with convective atmospheric flows \citep{chowdhuri2020revisiting}. We conducted an additional test by investigating whether the temporal temperature variances were dependent on the locations along the cable sections, and similar to PDFs and auto correlations, those did not show any spatial variations either (not shown). 

For statistical stationarity, the temporal auto correlations should not depend on the time origin ($t_0$) and must only be a function of temporal lags \citep{thomson2014data}. Corresponding to each of the five sections, we computed the temporal auto correlations of temperature fluctuations by changing the origin and the curves collapsed onto each other without showing any dependence on $t_0$. This result is showed in Fig. S3 of the Supplementary Material. Of course, there are more advanced tests available for checking stationarity in atmospheric flows \citep{pan2017determining} but for the present study this was sufficient. Therefore, we confirmed statistical homogeneity and stationarity for our clearcut flow situation. This outcome points towards a possibility that our measurement height (3.1 m) might be at the limit of the blending height during these buoyancy-dominated periods. 

\subsubsection{EM1 method}
\begin{figure*}[h]
\centering
\includegraphics[width=1\textwidth]{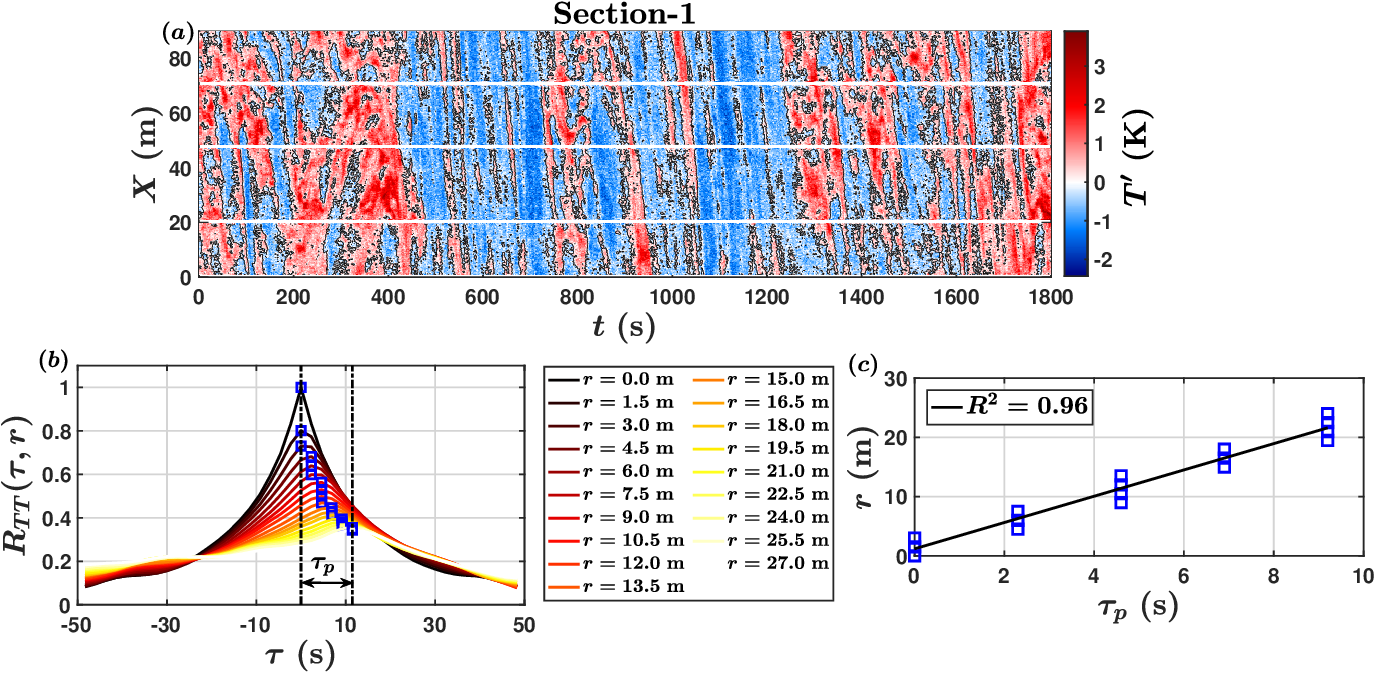}
  \caption{(a) The contour plot of temperature fluctuations ($T^{\prime}$) is shown from the DTS measurements of a specific section (Section-1) for a particular 30-min period between 10:30-11:00 UTC on 18 May 2024. The spatial locations along the cable section are denoted by $X$ and the temperature fluctuations are obtained by linearly detrending the temperature signals measured along time. The white horizontal strips indicate the regions where the DTS cable was fastened to masts and the temperature values were omitted from those locations. (b) The space-time correlation curves of temperature fluctuations ($R_{TT}(\tau,r)$) are shown for the same 30-min period obtained from Section-1. The temporal lags or leads are denoted as $\tau$ while the colors indicate different values of spatial separations along $X$, $r$. The blue squares denote the peaks of the correlation curves ($\tau_p$) for each value of $r$. (c) A scatter plot between $r$ and $\tau_p$ is shown whose slope represents $(U_e^2+V^2)/U_e$.}
\label{fig:3}
\end{figure*}
To compute the two parameters of the elliptic model, namely the convective and sweeping speeds ($U_e$ and $V$), one utilizes the peak values of space-time correlation curves by evaluating them in both temporal and spatial domains. To distinguish this method from the alternate one, we refer to it as EM1. These peak values are denoted as $\tau_p$ and $r_p$, respectively. Here, the $\tau_p$ values are associated with the peaks of the correlation curves evaluated over the temporal domain at different spatial separations. On the other hand, $r_p$ values are associated with the peaks of the correlation curves evaluated over the spatial domain at different instances in time. The elliptic form of space-time correlations suggests that the $\tau_p$ and $r_p$ follow a linear relationship with $r$ and $\tau$, respectively \citep{he2010small,zhou2011experimental}. As shown by \citet{zhou2011experimental}, these relationships can be written as,
\begin{equation}
\begin{split}
r_p=U_e \tau \\
\tau_p=\frac{U_e}{U_e^2+V^2} r.
\end{split}
\label{em1}
\end{equation}

To illustrate the applicability of the EM1 method on our DTS dataset, out of all the selected runs, for a specific representative 30-min period (10:30-11:00 UTC, 18 May 2024), we show the space-time contour plot of temperature fluctuations from Section 1 of the DTS cable (Fig. \ref{fig:3}a). The temperature values were omitted from those locations where the DTS cable was fastened to wooden structures. This decision was undertaken to reduce any positive bias in the temperature signal since these wooden structures were not entirely thermally inert. Corresponding to each cable section, these omitted values represented around 6\% of the cable length. Nevertheless, to ensure continuity in the data, those missing values were replaced with linearly-interpolated ones. Thereafter, the temperature fluctuations ($T^{\prime}$) were computed by linearly detrending the data over time separately at each location along the cable sections. The distance along Section 1 is denoted as $X$. To compute the convective speed of the temperature structures, we construct the space-time correlation curves from the DTS dataset. Mathematically, the two-point correlation functions of temperature fluctuations can be defined as,
\begin{equation}
    R_{TT}(\tau,r)=\frac{\overline{T^{\prime}(X,t)T^{\prime}(X+r,t+\tau)}}{\sigma_{T(X)}\sigma_{T(X+r)}},
    \label{st_corr}
\end{equation}
where $\tau$ are the temporal lags or leads, $r$ denotes the spatial separations along $X$, and $\sigma_{T(X)}$, $\sigma_{T(X+r)}$ are the standard deviations of temporal temperature fluctuations at locations $X$ and $X+r$. Henceforth, we use the notations $R_{TT}(\tau,r)$ and $R_{TT}(r,\tau)$ to differentiate between temporal and spatial cross-correlations of the DTS-derived temperature field, respectively.

For the same data as shown in Fig. \ref{fig:3}a, the quantity $R_{TT}(\tau,r)$ are plotted in Fig. \ref{fig:3}b. Here, different colors represent the increasing values of $r$ while the blue squares indicate the peaks ($\tau_p$) attained by $R_{TT}(\tau,r)$ for a specific $r$. By fitting a straight line to a scatter plot between $r$ and $\tau_p$, the slope would represent $(U_e^2+V^2)/U_e$ (see the solid black line in Fig. \ref{fig:3}c). To do this fit, we used the Matlab's fitlm function that has features to turn on the robustfit option (minimizes the impact of outliers) and set the intercept to 0. The goodness of the fit can be adjudged by its $R^2$ value, which was well-beyond 0.9 in this example case (Fig. \ref{fig:3}c). We restrict the $r$ values within 21 m while making the scatter plot, since for $r>21 \rm \ m$, the $R_{TT}(\tau,r)$ values are small and therefore the estimations of $\tau_p$ are not statistically robust. To ensure whether the shapes of $R_{TT}(\tau,r)$ curves were affected by the linear-interpolation operation, we increased the missing values to as large as 20\% of the cable length and replotted Fig. \ref{fig:3}b for this artificial data. We show this in Fig. S4 of the Supplementary Material and one could see that the peaks of $R_{TT}(\tau,r)$ curves behaved analogously as in the original data. 

\begin{figure*}[h]
\centering
\includegraphics[width=1\textwidth]{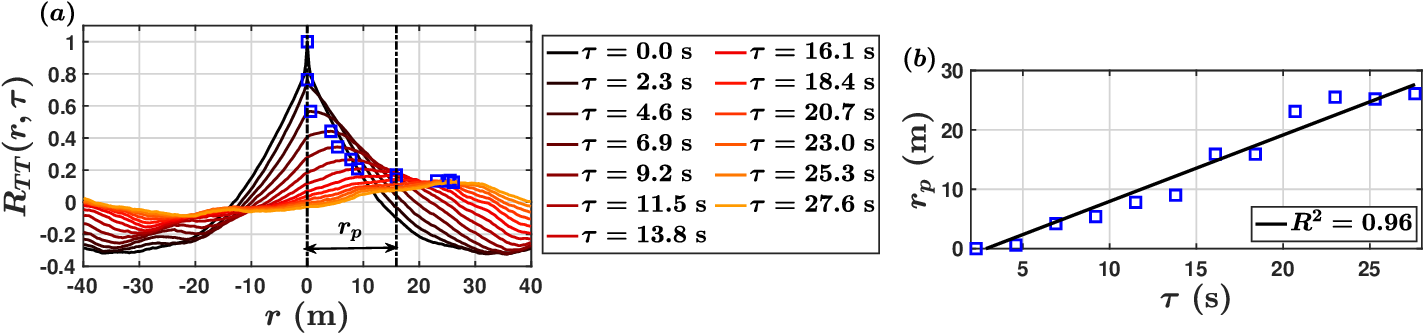}
  \caption{(a) Contrary to Fig. \ref{fig:3}b, the space-time correlation curves of temperature fluctuations ($R_{TT}(r,\tau)$) are shown where the $x$-axis represents the lags or leads in the spatial domain ($r$) while the colors indicate different values of temporal separations, $\tau$. The blue squares denote the peaks of the correlation curves ($r_p$) for each value of $\tau$. (b) A scatter plot between $r_p$ and $\tau$ is shown. The thick black color denotes the best fit line to compute $U_e$.}
\label{fig:4}
\end{figure*}

After estimating the slope of $\tau_p$-$r$ relationship, one does the same with $r_p$-$\tau$ one, which can be obtained by plotting the space-time correlation curves with respect to spatial lags or leads ($r$) at different values of temporal separations ($\tau$). For the same 30-min period akin to Fig. \ref{fig:3}b, we plot the $R_{TT}(r,\tau)$ curves in Fig. \ref{fig:4}a where the $x$-axis denotes the $r$ values while the colors of the curves indicate the values of $\tau$. Here, the spatial fluctuations were computed after removing a linear spatial trend. The spatial peaks of $R_{TT}(r,\tau)$ curves are denoted as $r_p$ (see the blue squares in Fig. \ref{fig:4}a). Similar to Fig. \ref{fig:3}c, one can do a scatter plot between $r_p$ and $\tau$ (see Fig. \ref{fig:4}b) to estimate the convective speed $U_e$. After computing $U_e$, one applies Eq. \ref{em1} to obtain the sweeping speed, $V$, as, 
\begin{equation}
    V=U_e\sqrt{\frac{1}{\beta_1 U_e}-1},
    \label{ss}
\end{equation}
where $\tau_p=\beta_1 r$.

\subsubsection{EM2 method}
\begin{figure*}[h]
\centering
\includegraphics[width=1\textwidth]{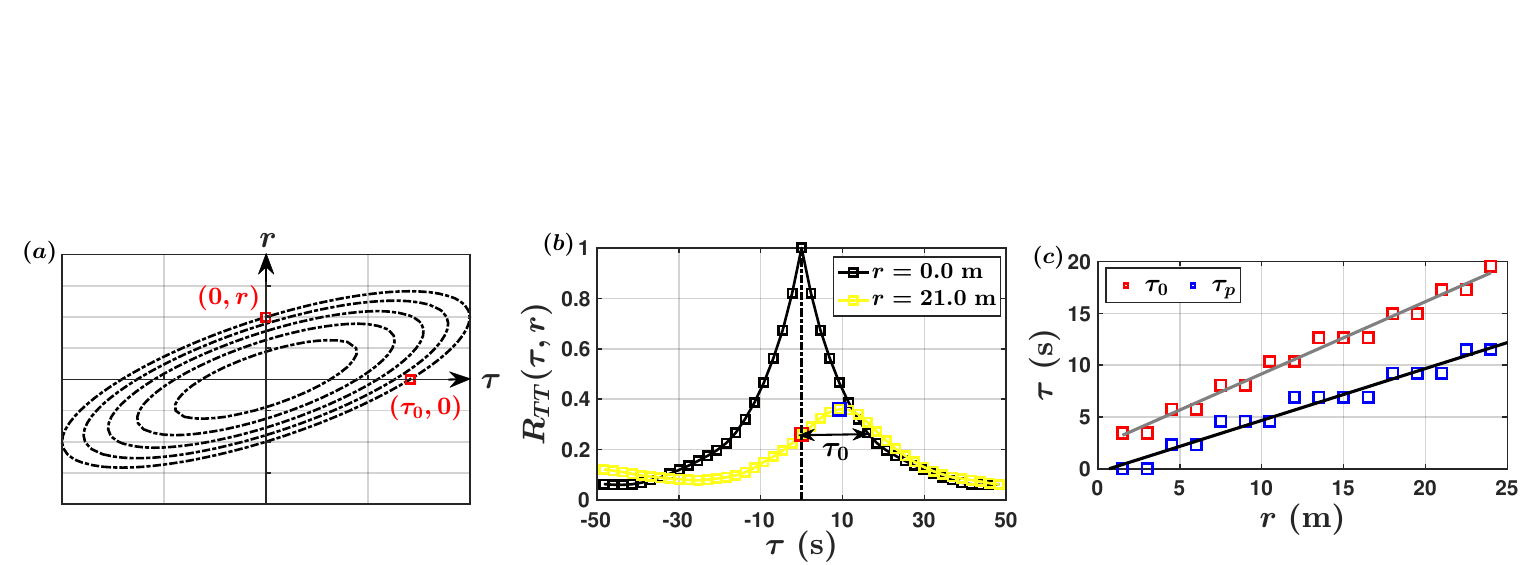}
  \caption{(a) A schematic of elliptic space-time iso-correlation contours are shown. The red squares denote two points on a specific iso-correlation contour with coordinates, $(\tau_0,0)$ and  $(0,r)$, respectively. (b) For the same 30-min period and Section-1 DTS measurements of Fig. \ref{fig:3}, $R_{TT}(\tau,r)$ curves are shown for two specific $r$ values in order to illustrate the computation of $\tau_0$. (c) The scatter plots between the separation distance $r$ and $\tau_0$ and $\tau_p$ are shown, where the slopes of the straight line fits are used to compute $\tilde{U_e}$ and $\tilde{V}$.}
\label{fig:5}
\end{figure*}

The EM2 method was developed to circumvent the need for having detailed spatial measurements to evaluate $R_{TT}(r,\tau)$ and estimate those parameters from $R_{TT}(\tau,r)$ itself \citep{he2015reynolds,he2017space}. Apart from that, the EM2 method proved to be useful in the case of very low wind speed. As argued by \citet{hogg2013reynolds}, for this specific situation, the peak positions of $R_{TT}(r,\tau)$ curves (i.e. $r_p$ values) with increasing $\tau$ values would not move since the mean wind speed is nearly zero. Yet, the sweeping effects will be present when the turbulence intensities are large and to compute them, one has to resort to the temporal domain. 

Similar to EM1, in EM2 method, one presumes beforehand that the space-time correlation contours are elliptic. For illustration purposes, in Fig. \ref{fig:5}a, we show a schematic diagram of the elliptic iso-correlation contours. Let us consider two points on a specific iso-correlation contour with coordinates $(\tau_0,0)$ and $(0,r)$ (denoted as red squares in Fig. \ref{fig:5}a). Since these two points lie on an iso-correlation contour, $R_{xx}(\tau_0,0)=R_{xx}(0,r)$ where $x$ could be any turbulent variable. By substituting the coordinates $(\tau_0,0)$ and $(0,r)$ in Eq. \ref{em2} and knowing that $R_{xx}(\tau_0,0)=R_{xx}(0,r)$, one gets the following expression,
\begin{equation}
\tau_0=\frac{1}{\sqrt{U_e^2+V^2}}r,
    \label{em2_1}
\end{equation}
which implies a linear relationship between $\tau_0$ and $r$ with a slope of $\beta_2$ where $\beta_2=1/\sqrt{U_e^2+V^2}$. By combining this information with $\tau_p=\beta_1 r$ where $\beta_1=U_e/(U_e^2+V^2)$ (see Eq. \ref{em1}), one obtains the following estimates of convective and sweeping speeds,
\begin{equation}
\begin{split}
\tilde U_e=\frac{\beta_1}{\beta_2^2} \\
\tilde V=\frac{1}{\beta_2}\sqrt{1-{\left(\frac{\beta_1}{\beta_2}\right)}^2},
\end{split}
\label{em2_2}
\end{equation}
where $\beta_1$ and $\beta_2$ are the slopes of the linear relationships between $\tau_p$-$r$ and $\tau_0$-$r$, respectively. In order to differentiate the estimations from EM1, a tilde sign is used on $U_e$ and $V$.

For the EM2 method to work, two time scales are required. One, the peak positions of $R_{TT}(\tau,r)$ curves ($\tau_p$), and second, the time scale $\tau_0$ at which $R_{TT}(\tau_0,0)=R_{TT}(0,r)$. This is demonstrated through Fig. \ref{fig:5}b, where the position of $\tau_p$ is shown by the blue square while $\tau_0$ is estimated as the temporal distance between the points where $R_{TT}(0,r)$ reaches its equivalent on the $R_{TT}(\tau,0)$ curve. In accordance with the EM2 method, a linear relationship was found between $\tau_0$-$r$ and $\tau_p$-$r$ (see Fig. \ref{fig:5}c). From the slopes of these linear relationships, the expressions in Eq. \ref{em2_2} were utilized to compute $\tilde U_e$ and $\tilde V$. As discussed before, we used the Matlab's fitlm function for computing the slopes.

To have robust estimates of $(U_e,V)$ and $(\tilde U_e,\tilde V)$, we restricted the periods when the associated $R^2$ values of $\tau_0$-$r$, $\tau_p$-$r$, and $r_p$-$\tau$ relationships exceeded 0.9. Although there were 113 half-hour periods when the direction of the mean wind was parallel to the cable sections 1 to 5, not for all such periods and for every cable section, statistically reliable estimates of $(U_e,V)$ and $(\tilde U_e,\tilde V)$ could be obtained. In Table \ref{tab:1}, we explicitly mention the number of data points corresponding to each section that were considered for the analysis. As one can see, the number of data points varied from one section to another. In total, there were 91 data points after consolidating the information from all the five sections. 

\begin{table}
\begin{center}
\begin{tabular}{|c |c |c |}
\hline
Section  & NOD & $d/h$\\
\hline
Section-1 & 11 & 0.9\\
\hline
Section-2 & 15 & 2 \\
\hline
Section-3 & 26 & 3.3\\
\hline
Section-4 & 17 & 5.1\\
\hline
Section-5 & 22 & 6.09\\
\hline
\end{tabular}
\caption{The number of data points (NOD) associated with individual parallel-wind sections for which one could obtain robust estimates of both $(U_e,V)$ and $(\tilde U_e,\tilde V)$. Here, $d$ refers to the distance of each cable section across the forest edge and $h$ denotes the averaged forest canopy height of 22.5 m (see Fig. \ref{fig:map}b).}
\label{tab:1}
\end{center}
\end{table}

\section{Results and discussion}
\label{results}
We begin with demonstrating the relationship between the convective speed estimations from the DTS measurements and the mean wind speed from the EC system at the clearcut while discussing the importance of the sweeping effects. To compute the convective and sweeping speeds, both EM1 and EM2 methods are employed and the results are compared between the two. The impact of these two methods on the collapse of space-time correlation curves and on their contour shapes are evaluated. To explain any difference between the two methods, a potential reason is presented. We end by showing that the random sweeping events influence the temperature measurements conducted by the EC system.

\subsection{Convective and sweeping speeds}
\begin{figure*}[h]
\hspace{-4cm}
\centering
\includegraphics[width=1\textwidth]{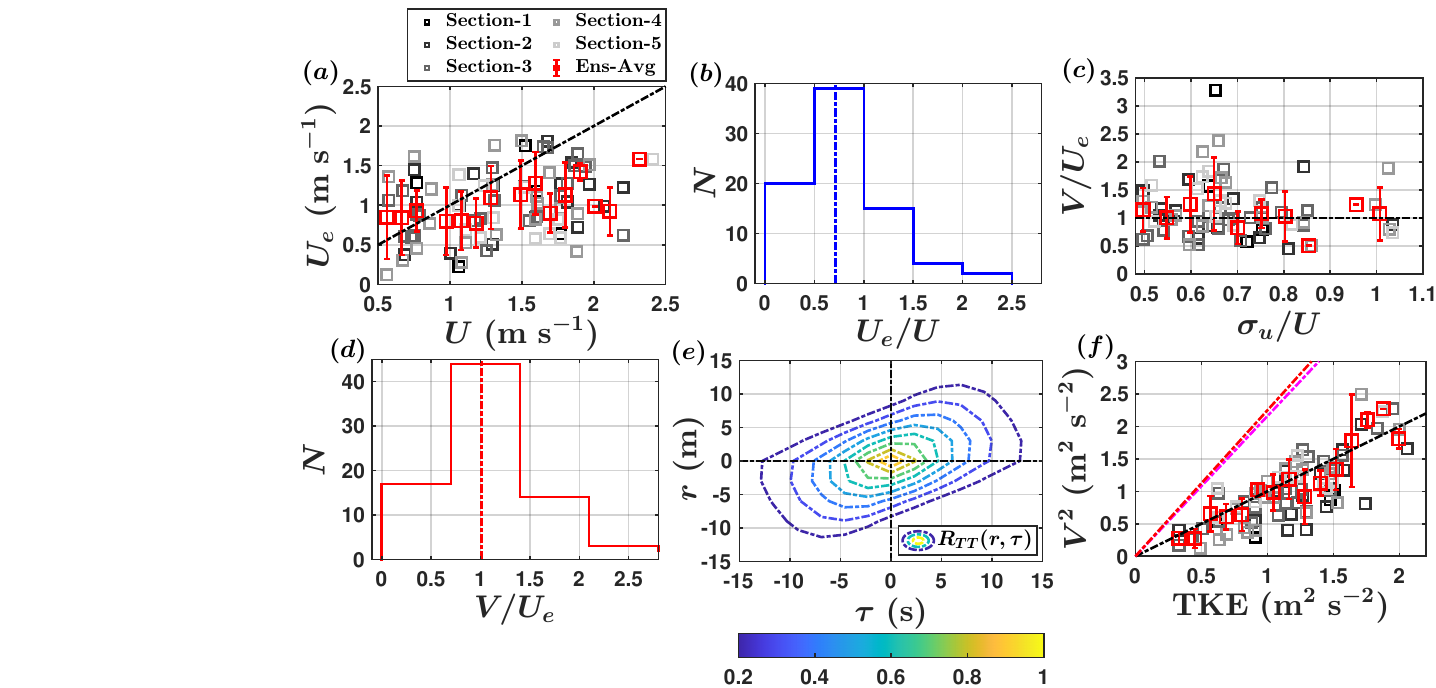}
\caption{For the cases when the direction of the mean wind was parallel to the forest edge, the scatter plots between (a) $U_e$ and $U$ are presented for the DTS cable sections 1 to 5. The different shades of gray squares indicate individual cable sections. The red squares with error-bars denote the bin averages with a spread of one standard deviation from the mean. Here, 20 linearly-spaced bins are formed after consolidating the data points from all the five cable sections. The dash-dotted black lines show the 1:1 relationship. (b) Regarding the same data, the histogram of $U_e/U$ is shown. The vertical dash-dotted lines indicate the median value of $U_e/U$. (c) The ratio between the sweeping and convective speeds ($V/U_e$) is plotted against the turbulence intensities ($\sigma_u/U$). Note that $U_e$ and $V$ are computed from the EM1 method. The red squares with error bars denote the bin averages and the spread. The horizontal dash-dotted black line represents $V/U_e=1$. (d) The histogram of $V/U_e$ is shown and the vertical dash-dotted line indicates the median value. (e) The ensemble-averaged contours of spatial cross-correlations ($R_{TT}(r,\tau)$) are shown for DTS cable section 1, whose ranges span from 0.2 to 0.9 at an increment of 0.1 and radiate outwards with the maximum value being concentrated at the innermost contour. (f) The relationship between the square of the sweeping speed ($V^2$) and turbulence kinetic energy (TKE) is shown for the DTS cable sections 1--5. The TKE is obtained from the EC system in the middle of the clearcut. The black, magenta, and red dash-dotted lines indicate the relationships $V^2 \approx \rm TKE$, $V^2 \approx \rm 2.16 \ TKE$, and $V^2 \approx \rm 2.25 \ TKE$, respectively. The latter two are obtained from \citet{everard2021sweeping} for flows over a vineyard canopy.}
\label{fig:6}
\end{figure*} 
Figure \ref{fig:6}a compares the convective speeds of the temperature structures estimated from the EM1 method ($U_e$) against the mean wind speed ($U$). We present these values for the conditions when the wind was blowing parallel to the forest edge. The gray squares in Fig. \ref{fig:6}a indicate the data points corresponding to sections 1 to 5 (see Table \ref{tab:1} for the number of data points). The red squares with error bars denote the bin-averaged values with the spread of one standard deviation. These bins are constructed after gathering the data points from all the five sections and then dividing them into 20 linearly spaced bins. Amid all the scatter, these bin-averaged values help to visualize the relationships better. From Fig. \ref{fig:6}a, one could notice that the $U_e$ values remain mostly smaller than $U$. By looking at the histogram of the ratio $U_e/U$, the median value appears to be at 0.7, shown as the blue vertical dash-dotted line in Fig. \ref{fig:6}b. Therefore, in a statistical sense, the $U_e$ values are nearly 0.7 times of $U$. 

In canonical wall-bounded turbulent flow and Rayleigh-B{\'e}nard convection experiments, depending on height and distance from the cell center, $U_e$ values of the velocity structures have been found to be lower than the mean wind speed \citep{zhou2011experimental,wang2014trpiv}. However, here we deal with convective speeds of temperature structures in a forest clearcut, for which no previous evidence exists. Thus, we compare our findings with \citet{everard2021sweeping} and \citet{han2022applicability,han2022predictive}, who reported such values for temperature in roughness sublayer and atmospheric surface layer flows. \citet{everard2021sweeping} observed that at heights within the vineyard canopy $U_e \approx 0.8U$ and at canopy top it was $U_e \approx 0.67U$ (see their Table 1 and Fig. 3c). Corresponding to an atmospheric surface layer flow over a homogeneous grassland, \citet{han2022applicability,han2022predictive} found $U_e$ was lower than $U$ for a wide range of stability conditions. They did not specify the relationship explicitly but from their figures it is clear that $U_e<U$. In contrast, \citet{su2000two}, by utilizing a large eddy simulation dataset, found the $U_e$ values for the velocity components and passive scalars to be larger than $U$ in a dense canopy flow. Regarding velocity, \citet{everard2021sweeping} provided an explanation for this disagreement by considering the fact that the correlation curves of temperature fluctuations behave differently than the velocity components and remain bounded between $u$ and $w$. Therefore, a one to one comparison between the convective speeds of velocity and temperature structures might not be appropriate. For our EC dataset, we find a similar behavior that the temperature auto-correlation curves are positioned in between of $u$ and $w$ components (not shown). Regarding the passive scalar part, in our case, the temperature fluctuations have more similarities with an active scalar (having non-Gaussian PDFs) and they belong from a clearcut rather than from a dense canopy. 


In general, depending on the flow conditions (canonical turbulent flows, roughness sublayer flows, atmospheric surface layer flows), surface characteristics (smooth, grassland, dense or sparse canopy, forest clearcut), and signal types (velocity and scalar), $U_e$ could be different from $U$. In the parlance of elliptic model, the geometry of the space-time correlation contours serve as a test for the validity of TH rather than the finding that $U_e \neq U$ \citep{he2006elliptic,zhao2009space,he2010small,wallace2014space,he2017space}. For instance, if one considers a scenario where $U_e \neq U$ and $V/U_e \ll 1$, one would find the space-time correlation contours to be nearly straight lines, in accordance with TH (see the schematic in Fig. \ref{fig:1}). However, when the sweeping speeds are not negligible and are of the same order as $U_e$, the space-time correlation contours deviate from straight lines and resemble closed-form curves. In fact, this situation occurs when the turbulence intensities of the flow are large \citep{zhao2009space}. We next highlight the importance of the sweeping effects for our flow conditions. 

\citet{flesch_wind_1999} observed highly turbulent wind conditions in a forest clearcut flow approximately at a distance of three to five times the canopy height. This distance matches with the location of our EC set up, situated nearly five times the canopy height from the forest edge. In Fig. \ref{fig:6}c, we plot the ratios of sweeping and convective speeds ($V/U_e$) against the streamwise turbulence intensities ($\sigma_u/U$) for our forest clearcut flow. These turbulence intensities are computed from the EC measurements at the middle of the clearcut. One could immediately notice that $\sigma_u/U$ for our flow situation is quite high with the minimum value being 0.48. The standard practice in micrometeorology often considers $\sigma_u/U>0.5$ as highly turbulent \citep{willis1976use}, so these $\sigma_u/U$ values were very near or beyond that limit.

For such a highly turbulent situation, as demonstrated in Fig. \ref{fig:6}c, the sweeping speeds are indeed of the same order as $U_e$ if not larger. This same information can be presented in terms of their histogram (Fig. \ref{fig:6}d) and it is clear that the histogram peaks around the value of $V/U_e=1$, which happens to be the median as well (see the red vertical dash-dotted line in Fig. \ref{fig:6}d). These strong sweeping effects introduce a finite amount of curvature to the space-time correlation contours and therefore, their shapes do not remain equivalent to a straight line. To demonstrate this through an example, in Fig. \ref{fig:6}e, the ensemble-averaged iso-correlation contours of $R_{TT}(r,\tau)$ are plotted for cable section 1. 

These contours in Fig. \ref{fig:6}e span between the range of 0.2 to 0.9 at an increment of 0.1 and radiate outwards with the largest values being associated with the innermost ones. It is abundantly clear that these contours are closed-form curves, unlike a family of straight lines as predicted by TH (see the schematic in Fig. \ref{fig:1}). Moreover, they remain symmetrically oriented along a direction that coincides with the convective speeds of the temperature structures. Since these contours display a conspicuous tilt, it is clear that TH is only valid as a first-order approximation. On the other hand, the aspect ratios of these contours (height divided by width) are determined by how strong the sweeping effects are in comparison to $U_e$ \citep{zhao2009space}. 

To verify whether these sweeping effects are caused by large scale eddies, as proposed by Kraichnan in his random sweeping hypothesis \citep{kraichnan1964kolmogorov}, one could study the scaling relationship between the square of the sweeping speeds ($V^2$) and the turbulence kinetic energy (TKE) of the flow. \citet{everard2021sweeping} showed that $V^2$ was intimately connected to the TKE of the roughness sublayer flow. They found this relationship to be $V^2 \approx \rm 2.16 \ TKE$ for heights above a vineyard canopy and $V^2 \approx \rm 2.25 \ TKE$ for heights within the canopy (see the red and magenta dash-dotted lines in Fig. \ref{fig:6}f). We plot the same $V^2$-TKE relationship in Fig. \ref{fig:6}f where the TKE is independently estimated from the EC system, defined as $(\sigma_u^2+\sigma_v^2+\sigma_w^2)/2$. 

Similar to \citet{everard2021sweeping}, we do see a strong relationship between the two. However, as opposed to them, the relationship in Fig. \ref{fig:6}f emerges to be $V^2 \approx \rm TKE$ (black dash-dotted line in Fig. \ref{fig:6}f). This discrepancy is consistent with the expectation that the boundary conditions on the flow, one of which is the underlying surface characteristics, could potentially alter the relationship between $V^2$ and TKE \citep{everard2021sweeping}. To compare, for the vineyard canopy flow studied by \citet{everard2021sweeping}, the averaged height of the roughness elements was nearly 2.3 m. For our clearcut flow, the height of the roughness elements varied between 0.5 to 1.5 m. 

Although the scaling relationship between $V^2$ and TKE was originally proposed for isotropic turbulence \citep{tennekes1975eulerian}, it is exceptional that the same holds for anisotropic turbulence. Specific to our flow situation, the turbulence was anisotropic at both small and large scales. Here, large scale anisotropy was determined through the invariants of the Reynolds stress tensor \citep{chowdhuri2020revisiting} while the small scale anisotropy was assessed through the intermittency-anisotropy framework of \citet{chowdhuri2024quantifying} (results not shown). Interestingly, the presence of large scale anisotropy explains why the ratios $V/U_e$ do not seem to increase with our range of $\sigma_u/U$ values, even when $V \propto ({\rm TKE})^{1/2}$ and $U_e \propto U$ (see Appendix B).

\subsection{Comparison between EM1 and EM2}
\label{conversion}
\begin{figure*}[h]
\centering
\includegraphics[width=1\textwidth]{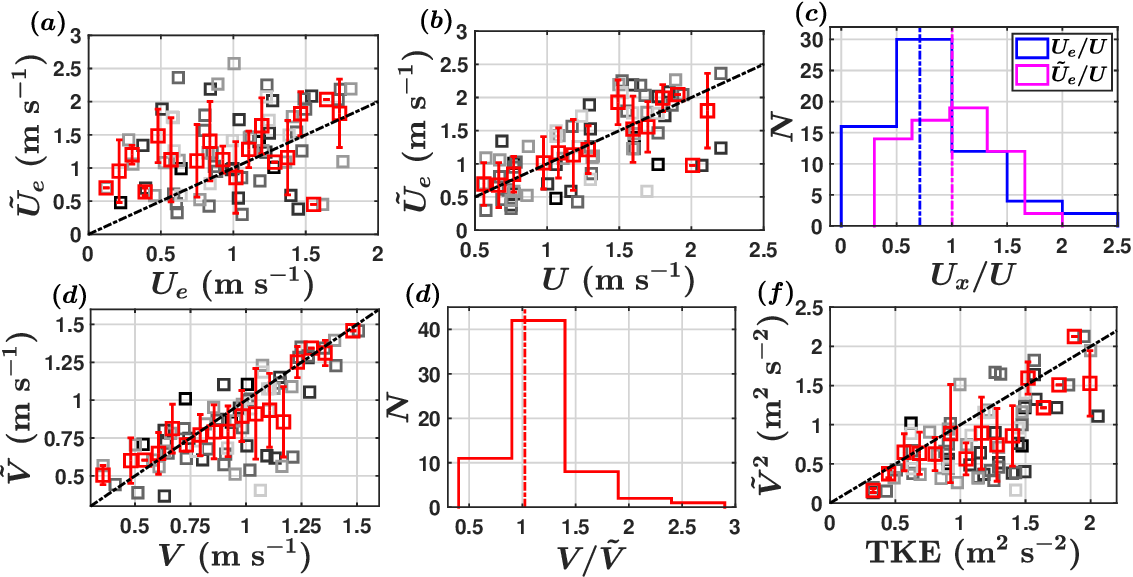}
\caption{For the same parallel-wind scenario, the scatter plots are presented between (a)$\tilde U_e$ and $U_e$ and (b) $\tilde U_e$ and $U$. The tilde sign indicates that the computations are performed by applying the EM2 method. The squares in the scatter plots carry similar information as in Fig. \ref{fig:6}. (c) The histograms are presented for the ratios $U_e/U$ and $\tilde U_e/U$ and the vertical dash-dotted lines denote their respective median values. (d)--(e) Comparisons between $\tilde V$ and $V$ are shown in the form of a scatter plot and a histogram. (f) The $\tilde V^2$-$\rm TKE$ relationship is shown but for the EM2 method.}
\label{fig:7}
\end{figure*}

The results presented in Fig. \ref{fig:6} show evidences that strong sweeping effects are present in our clearcut flow. Note that the computations of $U_e$ and $V$ in Fig. \ref{fig:6} are done by EM1 method. However, as discussed previously, there exists an alternate method, namely EM2 to compute the same. The EM2 method was applied by \citet{han2022applicability} for an atmospheric surface layer flow by utilizing the same DTS dataset of \citet{cheng2017failure}. They found that differences smaller than 20\% existed between the estimations from the two methods. They could not provide any reason behind this difference and eventually used an averaged value of $U_e$ and $V$ obtained from the two methods. For the forest clearcut flow analyzed here, we investigate by how much the estimations from the EM1 and EM2 methods vary with respect to each other.

To do so, in Fig. \ref{fig:7}, we present comparisons between the estimates of $(U_e,V)$ and $(\tilde U_e,\tilde V)$. The tilde sign indicates that the estimates have been obtained from the EM2 method. From the scatter plot between $U_e$ and $\tilde U_e$, it is clear that the $\tilde U_e$ values are mostly larger than $U_e$ (Fig. \ref{fig:7}a). In fact, upon comparing $\tilde U_e$ with $U$, it appears that the $\tilde U_e$ values are approximately equal to the mean wind speed (Fig. \ref{fig:7}b). This is evident from the histograms of $\tilde U_e/U$ and $U_e/U$, where the median value of $\tilde U_e/U$ happens to be nearly equal to unity rather than 0.7 as obtained for $U_e/U$ (Fig. \ref{fig:7}c). On the other hand, by comparing $\tilde V$ and $V$, they seem to lie on the 1:1 line (Fig. \ref{fig:7}d). By inspecting the histogram of $V/\tilde V$, the median value of $V/\tilde V$ is indeed 1 (see the red dash-dotted vertical line in Fig. \ref{fig:7}e). Moreover, the $\tilde V^2$-TKE scaling relationship in Fig. \ref{fig:7}f is not very different from the $V^2$-TKE one in Fig. \ref{fig:6}f, albeit with a somewhat more scatter.

This comparison presents a compelling outcome. From the EM2 method, one obtains the convective speeds of the temperature structures to be almost equal to the mean wind speed, while the same appears to be lower when the EM1 method is applied. This difference is clearly more than 20\% as found by \citet{han2022applicability} for an atmospheric surface layer flow. Irrespective of the methods, the estimates of the sweeping speeds do not differ much. We do present a thorough investigation on this aspect, but first we ask if this disparity could be caused by noise in the DTS data.

\subsection{Influence of noise}
\begin{figure*}[h]
\centering
\includegraphics[width=1\textwidth]{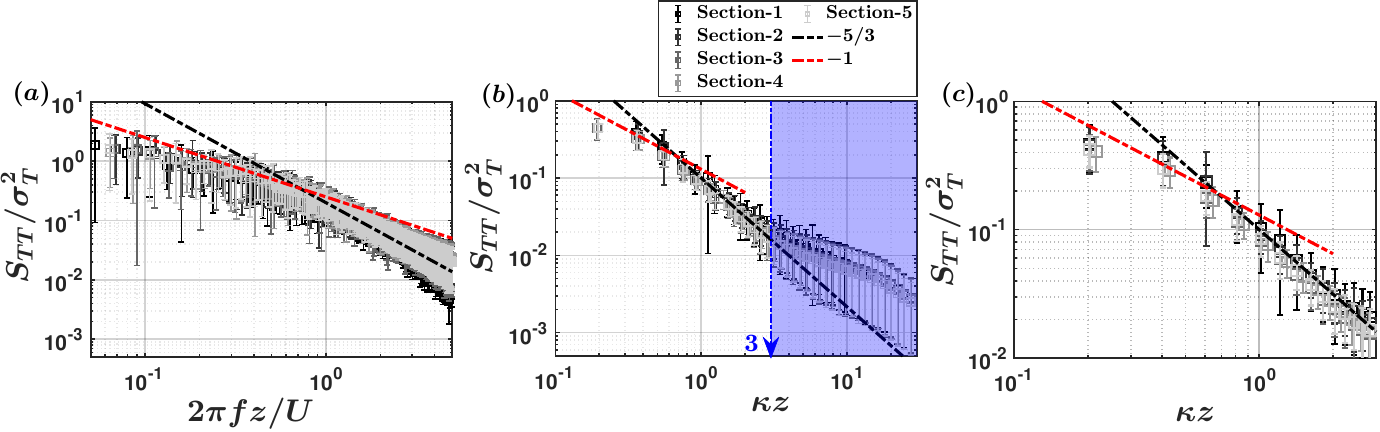}
  \caption{The ensemble-averaged DTS temperature spectra are shown for (a) temporal and (b) spatial domains, respectively. The error bars signify one standard deviation from the ensemble mean. The parallel-wind DTS cable sections (sections 1-5) are used for illustration purposes. The red and black dash dotted lines in (a) and (b) represent the $-1$ and $-5/3$ spectral slopes, respectively. The blue-shaded region in (b) is associated with those higher wavenumbers ($\kappa z > 3$) where the DTS spatial spectra have a slope shallower than the expected $-5/3$ and have been removed through a spatial-smoothing operation. (c) The spatially-smoothed DTS spectra are shown at a resolution of 3 m after averaging 10 consecutive spatial points.}
\label{fig:8}
\end{figure*}
Like any other observations, the DTS measurements are not perfect and are to a certain degree contaminated by noise and possibly also other instrument artifacts \citep{peltola2021suitability}. Usually, the noise is related to the spectral features of DTS data \citep{cheng2017failure,peltola2021suitability}. For the parallel-wind DTS cable sections, in Fig. \ref{fig:8}, we show the spectra of temperature fluctuations computed over the temporal (Fig. \ref{fig:8}a) and spatial domains (Fig. \ref{fig:8}b), respectively. The error bars indicate one standard deviation with respect to the ensemble mean and the varying shades of gray represent different sections. The spectral amplitudes ($S_{TT}$) of individual sections are normalized by the respective temperature variances ($\sigma^2_{T}$) and the frequencies are scaled with the standard surface layer scaling ($2\pi fz/U$) of \citet{kaimal1972spectral}, where $U$ is the mean wind speed from the EC system. 

From Fig. \ref{fig:8}a, one could see that the DTS temperature spectra in the temporal domain follow a $-5/3$ slope at smaller scales while displaying a $-1$ slope at larger scales. The $-1$ spectral slope in temperature has been noted before for mildly-unstable atmospheric surface layer flows over a lake and a grassland \citep{li2016k}. Here, we deal with roughness sublayer flows over a highly-heterogeneous surface and therefore the results cannot be directly compared. Moreover, the temperature signal in our case does not behave like a passive scalar as our periods are selected from convective conditions. It is thus not straightforward to associate this $-1$ spectral slope directly with attached eddies commonly found in near-neutral flows. We also observe a break in the $-1$ spectral slope at $2\pi fz/U <0.2$, which possibly indicates the presence of boundary-layer scale eddies. To see these features more clearly, in Fig. S5 of the Supplementary Material, we present examples of the temporal spectra only from sections 1 and 5 without the error bars. Upon comparison of DTS temporal spectra with the along-wind EC temperature spectra, a satisfactory agreement was found between the two (not shown). However, since the temporal resolution of DTS data ($2.3 \ \rm s$) was lower than the EC one ($0.1 \ \rm s$), it did not fully capture the location of the spectral peak. 

On the other hand, for the spatial domain, the DTS temperature spectra show a rather narrow region of $-1$ spectral slope (Fig. \ref{fig:8}b). Here, $\kappa$ are the streamwise wavenumbers computed from the spatial DTS data and scaled with the measurement height, $z$. Since the DTS cable length at each section was nearly 100 m, it did not capture enough large scale eddies to show an extended $-1$ slope as in the temporal spectra. At smaller scales, the DTS spatial temperature spectra displayed a $-5/3$ slope up to wavenumbers $\kappa z \leq 3$, followed by a region with a spectral slope shallower than $-5/3$. This region, $\kappa z> 3$, is shaded with blue color in Fig. \ref{fig:8}b.

\citet{cheng2017failure} observed a similar feature in their DTS data and termed this enhanced blue-shaded region in Fig. \ref{fig:8}b as noise (see their Fig. 2c). Apart from \citet{cheng2017failure}, there is a dearth of previous studies where spatial spectra of temperature have been reported from near surface observations. The large eddy simulation study of \citet{tong2015multipoint} reported a $-5/3$ region in the spatial spectra of temperature at heights within the convective surface layer. Unlike temporal domain, in the absence of other independent measurements to confirm the spatial spectra, we decided to discard the blue-shaded region by applying a smoothing filter. The limit $\kappa z=3$ is nearly equivalent to a wavelength of 6 m, which corresponds to a spatial separation of 3 m. Therefore, to truncate the spectra beyond $\kappa z>3$, we averaged 10 consecutive spatial points, which artificially reduced the spatial resolution from 30 cm to 3 m. For this artificially reduced data, we computed the spatial DTS spectra and as expected, the spectra of this smoothed dataset only revealed the $-5/3$ spectral slope (Fig. \ref{fig:8}c).

\begin{figure*}[h]
\centering
\includegraphics[width=1\textwidth]{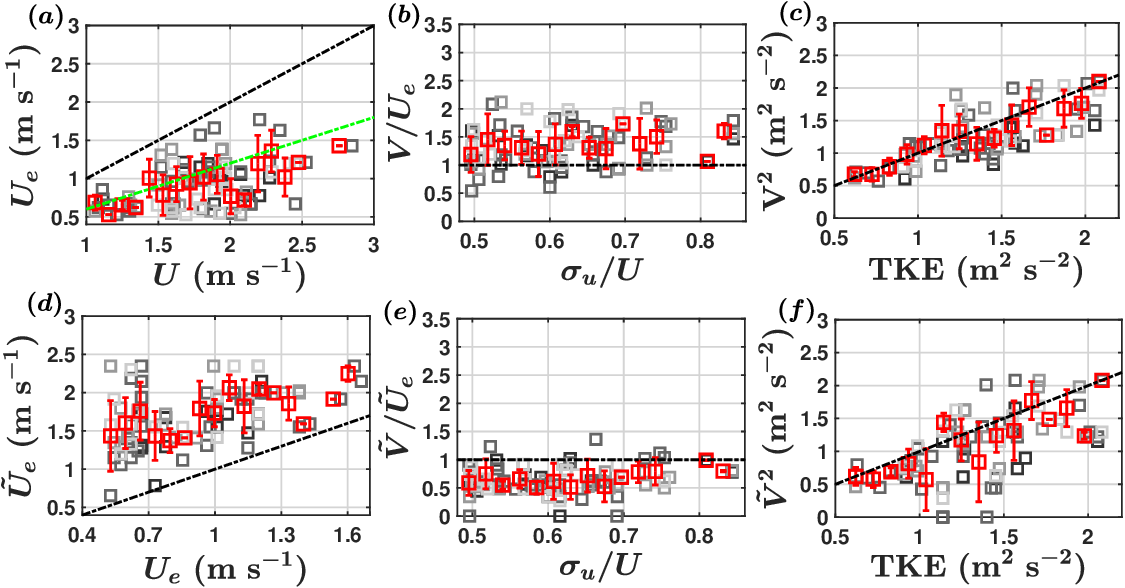}
\caption{(a)--(f) The EM1 and EM2 comparisons between the elliptic model parameters and their scaling relationships with the turbulence kinetic energy are presented for the spatially-smoothed 3-m resolution DTS temperature data. The green dash-dotted line in (a) indicates the relationship $U_e=0.7U$ while the black dash-dotted lines in (a), (c), (d), and (f) indicate the 1:1 relationship. The horizontal black dash dotted lines in (b) and (e) signify $V/U_e=1$ and $\tilde V/\tilde U_e=1$, respectively.}
\label{fig:9}
\end{figure*}

After removing this noise, it is imperative to check whether our estimations of the elliptic model parameters ($U_e$ and $V$) are in any way affected by this operation. In Fig. \ref{fig:9}, corresponding to the 3-m spatial resolution DTS temperature data, we present the estimations of $U_e$ and $V$ from both EM1 and EM2 methods. It is clear that their statistical characteristics are not impacted by the smoothing operation. For instance, upon comparing $U_e$ with $U$, we find that the $U_e$ values remain smaller than $U$ and the relationship between them is $U_e \approx 0.7U$ (Fig. \ref{fig:9}a). This is shown as green dash-dotted line in Fig. \ref{fig:9}a, fitted to the red squares representing the bin-averaged values. A similar outcome for the original DTS data (30-cm resolution) was observed in Fig. \ref{fig:6}b, but with a larger scatter. 

Apart from that, the sweeping speeds remain of the same order as $U_e$ and its squared values follow a 1:1 relationship with TKE (Figs. \ref{fig:9}b--c). To confirm further, we compare the $R_{TT}(\tau,r)$ and $R_{TT}(r,\tau)$ curves between the spatially-smoothed and original DTS temperature data for a specific 30-min period and their shapes do not appear to be different either (see Fig. S6 in the Supplementary Material). Even regarding the EM2 method, the findings are the same as in Fig. \ref{fig:7}, where $\tilde U_e$ is larger than $U_e$ and $\tilde V^2 \approx \rm TKE$ (Figs. \ref{fig:9}d and f). However, upon closely observing the individual gray-shaded data points in Fig. \ref{fig:9}f and given the $y$-axis limit is the same in both Figs. \ref{fig:9}c and f, one could infer that the $\tilde V$ values are, to a certain extent, lower than $V$. This is one aspect that was not very clear in Figs. \ref{fig:7}a and d, since the relationships between $\tilde U_e$-$U_e$ and $\tilde V_e$-$V$ displayed more scatter. One point to consider here is that for the EM2 method, the ratio $\tilde V/\tilde U_e$ remains lower than unity despite having large turbulence intensities (Fig. \ref{fig:9}e). This occurs because the EM2 method yields larger values of the convective speeds than EM1 while the sweeping speed estimates remain slightly smaller. Eventually, the larger (smaller) $\tilde U_e$ ($\tilde V$) values reduce the ratio $\tilde V/\tilde U_e$. In a nutshell, the results in Fig. \ref{fig:9} increase our confidence in the estimations of these bulk parameters, since their statistical characteristics remain robust with respect to the high wavenumber noise in the DTS data. 

\subsection{Critical assessment of elliptic model}
Hitherto, we have demonstrated that significant sweeping effects associated with large scale eddies persist for our highly turbulent flow conditions ($\sigma_u/U \geq 0.48$) in a forest clearcut. To account for these random sweeping effects, an elliptic model might be more appropriate to describe the relationship between space and time. However, we find that two methods, namely EM1 and EM2, yield different results for the elliptic model parameters, and the outcome remains robust with respect to the high wavenumber noise in the DTS data. This disparity is more prevalent with respect to $U_e$ but not so much for $V$. 

Both EM1 and EM2 methods are empirical and assume that the space-time correlation contours are elliptic, while in reality their morphology could be different. Therefore, as shown by previous research, the validity of the elliptic model parameters should be assessed by conducting two tests \citep{he2006elliptic,zhao2009space,he2010small,zhou2011experimental,he2014space,han2022predictive}. First, to see if the elliptic model parameters can collapse the temporal cross-correlation curves at few spatial locations when the temporal lags are converted to spatial scales through an elliptic scaling. Second, to check if the shapes of the space-time iso-correlation contours can be expressed as ellipses by using the parameters of the elliptic model. We present results from these two tests by using both EM1 and EM2 methods. The objective is to evaluate which one of the two methods explains the properties of the space-time correlations the best.

\subsubsection{Space-time correlations}
\begin{figure*}[h]
\centering
\includegraphics[width=1\textwidth]{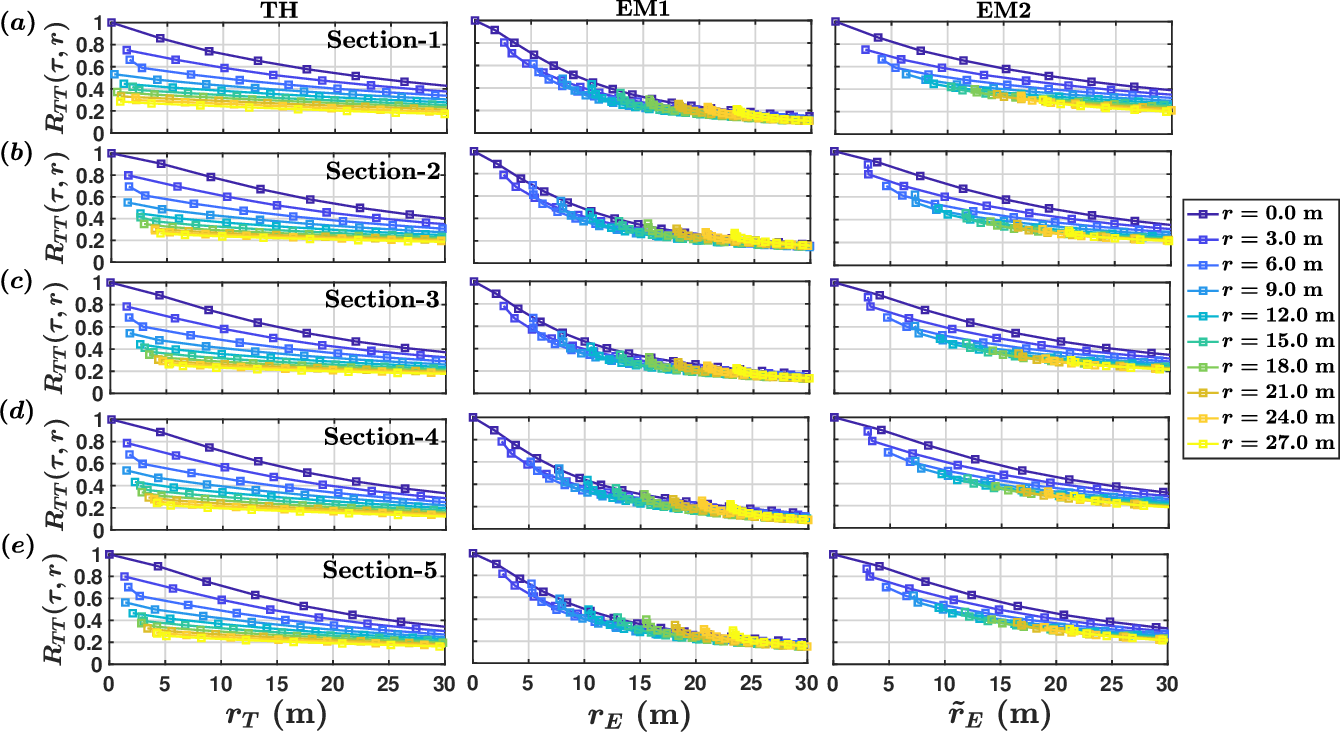}
  \caption{(a)--(e) For the spatially-smoothed DTS measurements at 3-m resolution, temporal cross-correlation curves of temperature fluctuations ($R_{TT}(\tau,r)$), computed between different streamwise spacings ($r$), are plotted. The colors represent these different streamwise spacings (see the legend). Each row corresponds to individual cable sections (sections 1 to 5) and the columns are associated with three methods being employed to convert the temporal lags ($\tau$) to spatial separations, which are TH ($r_T$), EM1 ($r_E$), and EM2 ($\tilde r_E$), respectively (see Eqs. \ref{es} and \ref{TH}).}
\label{fig:10}
\end{figure*}
The elliptic scaling dictates that the temporal cross-correlation curves of any turbulent variable (e.g., $R_{TT}(\tau,r)$), computed between points separated in space (i.e. between locations $X$ and $X+r$, where $r$ is the separation distance), would collapse to a single curve when the temporal lags ($\tau$) are converted to spatial separations ($r_E$) using,
\begin{equation}
    r_E=\sqrt{(r-U_e\tau)^2+V^2\tau^2},
    \label{es}
\end{equation}
where $U_e$ and $V$ are the convective and sweeping speeds, respectively \citep{zhao2009space,zhou2011experimental}. The $r_E$ scaling produces a perfect collapse of the temporal cross-correlation curves when the space-time contours are elliptic. This scaling contrasts with the linear scaling of TH, which dictates,
\begin{equation}
    r_T=\vert r-U\tau \vert,
    \label{TH}
\end{equation}
where $U$ is the mean wind speed. Since we keep our $\tau$ values as positive quantities, the modulus sign is inserted in Eq. \ref{TH} to avoid $r_T$ being negative. Note that one could have used $U_e$ in Eq. \ref{TH} as well and that would not have impacted the scaling results. Regarding the elliptic scaling, we compute them by using both $(U_e,V)$ and $(\tilde U_e,\tilde V)$ in Eq. \ref{es}, and refer to the obtained values as, $r_E$ and $\tilde r_E$, respectively.

Figure \ref{fig:10} shows the temporal cross-correlation curves ($R_{TT}(\tau,r)$) for the parallel-wind scenario, corresponding to five different sections by utilizing the spatially smoothed DTS data (3-m resolution). The different colored lines indicate the $r$ values spanning up to 27 m. The horizontal five rows are associated with individual cable sections. The three columns represent the $R_{TT}(\tau,r)$ curves where the temporal lags ($\tau$) are converted to spatial separations, either by using TH ($r_T$ from Eq. \ref{TH}), or by EM1 and EM2 methods ($r_E$ and $\tilde r_E$ from Eq. \ref{es}). For each section, the $R_{TT}(\tau,r)$ curves have been ensemble averaged over the number of runs as shown in Table \ref{tab:1}. To compute $r_T$, the mean wind speed from the EC system is used. For $r_E$ and $\tilde r_E$, we use the section-specific values of $(U_e,V)$ and $(\tilde U_e,\tilde V)$.  Instead of considering the whole range of spatial scales, we restrict them up to 30 m, beyond which the $R_{TT}(\tau,r)$ values drop significantly below $1/\exp(1)$. For a better visualization, we use a linear $x$-axis in Fig. \ref{fig:10}. 

From Fig. \ref{fig:10}, it is apparent that the TH method performs poorly to reduce the spread between the space-time correlation curves at different $r$ values (see the family of curves corresponding to the first column of Fig. \ref{fig:10}). On the other hand, irrespective of the cable sections, the spread reduces dramatically when the EM1 scaling is applied (see the family of curves corresponding to the middle column of Fig. \ref{fig:10}). In the absence of sweeping effects ($V \to 0$), both $r_T$ and $r_E$ scaling are linear in nature (see Eqs. \ref{es} and \ref{TH}) and should have identical impact on the collapse of the space-time correlation curves. Therefore, this discrepancy between the two indicates that the sweeping effects influence the space-time relationship at all the $r$ ranges considered here. The EM2 method definitely performs better than TH but the collapse of those curves are not as tight as in EM1 (third column of Fig. \ref{fig:10}). Note that although the $r_E$ scaling could bring the temporal cross-correlation curves close together, the collapse was not a perfect one as commonly found in laboratory experiments \citep{zhou2011experimental}.

\subsubsection{Space-time contours}
\label{em_diff}
\begin{figure*}[h]
\centering
\hspace{-2cm}
\includegraphics[width=1\textwidth]{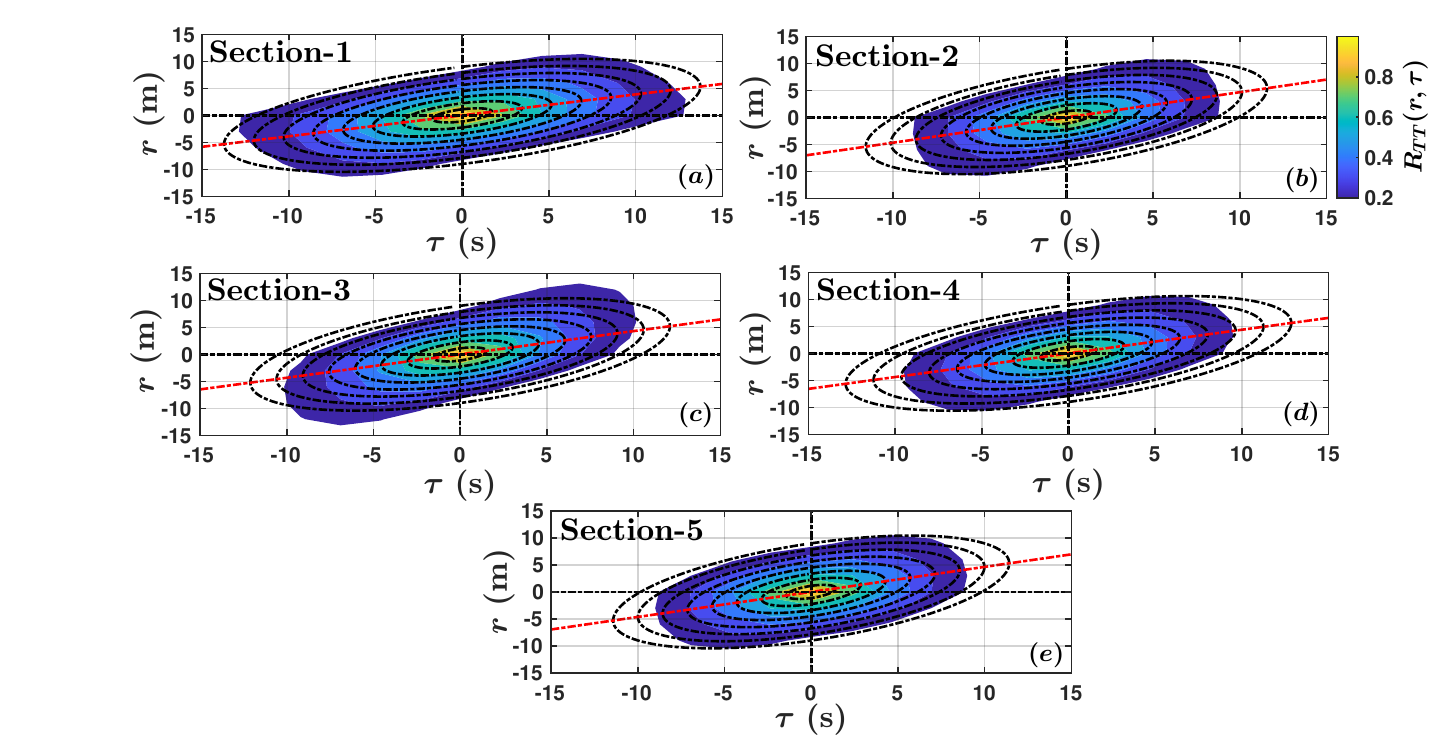}
\caption{(a)--(e) The empirically determined space-time temperature iso-correlation contours ($R_{TT}(r,\tau)$) from the spatially smoothed DTS data are compared against the elliptic model for all the five DTS cable sections. In (a) to (e), the empirically determined contours are shown as filled ones, whose ranges span from 0.2 to 0.9 at an increment of 0.1 and radiate outwards with the maximum value being concentrated at the innermost contour. The black dash-dotted lines correspond to ellipses of distinct $r_E$ values that vary from 0 to 9 m at an increment of 1.125 m. To compute these black dash-dotted ellipses, the EM1 parameters are used. The red dash-dotted lines in (a)--(e) denote the fit $r=U_e\tau$, where $U_e$ is the convective speeds of the temperature structures estimated from the EM1 method.}
\label{fig:12}
\end{figure*}
We now investigate the second aspect, i.e. if the space-time iso-correlation contours can be expressed as ellipses by using the elliptic model parameters. To do so, we compute the space-time correlation contours by evaluating the spatial cross-correlations of temperature between different instances in time. We conduct this exercise in the spatial domain because the temporal cross-correlation curves are not oriented along a direction that coincides with the convective speeds of the temperature structures but in a direction that is determined by the slope of the $r$-$\tau_p$ relationship (Eq. \ref{em1}), i.e. $(U_e^2+V^2)/U_e$ (see Fig. S7 in the Supplementary Material where an example is presented from section 1). These empirically determined contours of spatial cross-correlations ($R_{TT}(r,\tau)$) are presented in Fig. \ref{fig:12} for all the five cable sections and they have all been ensemble averaged. These iso-correlation contours of $R_{TT}(r,\tau)$ are plotted in the range of 0.2 to 0.9 at an increment of 0.1 and are shown as the filled ones. The orientation of these contours is represented by a red dash-dotted line in Fig. \ref{fig:12}, which denotes $r=U_e\tau$ where $U_e$ values are ensemble-averaged and determined from the EM1 method.

\begin{figure*}[h]
\centering
\hspace{-2cm}
\includegraphics[width=1\textwidth]{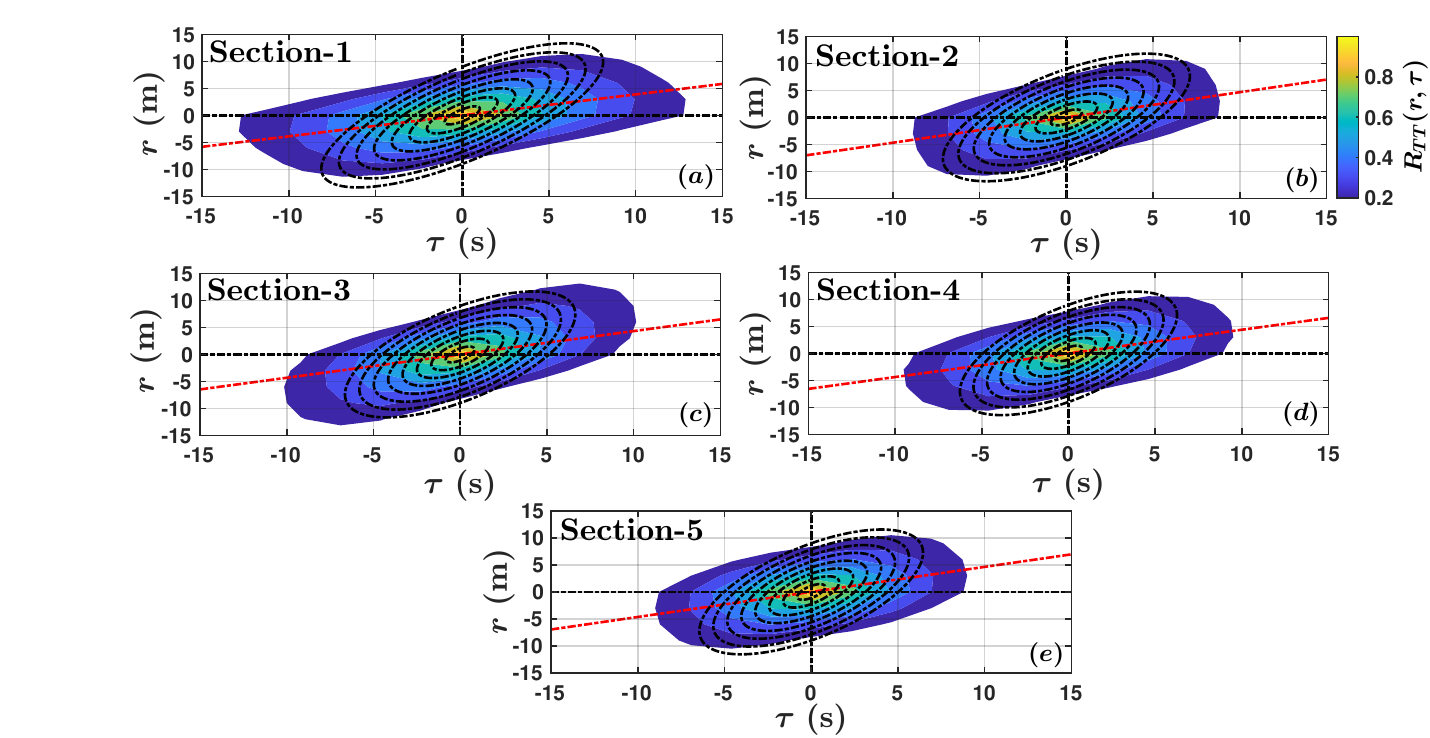}
\caption{Same as in Fig. \ref{fig:12}, but the black dash-dotted ellipses are computed by using the EM2 parameters.}
\label{fig:13}
\end{figure*}

The elliptic model suggests that each iso-correlation contour of $R_{TT}(r,\tau)$ can be associated with a unique $r_E$ value that relates $r$ and $\tau$ through an ellipse (see Eqs. \ref{em2} and \ref{es}). To determine how well the morphology of $R_{TT}(r,\tau)$ iso-correlation contours match with elliptic shapes, one could essentially overlay a family of ellipses at different $r_E$ values on these space-time contours. While doing this fitting exercise, one could have chosen the $r_E$ values such that they match with the exact iso-correlation levels as in the observations but that requires knowing the Taylor microscale (see Eq. \ref{em2}), which cannot be estimated either from DTS or EC measurements.  Therefore, to conduct this exercise, we chose a range of $r_E$ values that varied between 0 to 9 m at an increment of 1.125 m. As our objective is to qualitatively evaluate how well the closed-form curves of $R_{TT}(r,\tau)$ contours can be expressed as ellipses, the exact details regarding the range of $r_E$ is irrelevant here. After deciding on the range, corresponding to each $r_E$ value, one could get an elliptic relationship between $r$ and $\tau$, expressed in the parametric form as, 
\begin{equation}
\begin{split}
\tau=\frac{r_E }{V}\sin(\theta) \\
r=r_E \cos(\theta)+\frac{U_e r_E }{V}\sin(\theta),
\end{split}
\label{es1}
\end{equation}
where, $0 \leq \theta \leq 2\pi$. Here, $U_e$ and $V$ are the convective and sweeping speeds, and we estimate those from the EM1 method. 

These parametric relationships at distinct $r_E$ values are shown as black dash-dotted ellipses in Fig. \ref{fig:12}. After noticing the agreement in shapes between the filled $R_{TT}(r,\tau)$ contours and black dash-dotted ellipses, it is clear that the EM1 method describes both the orientation and aspect ratios of the space-time correlation contours quite well. In other words, the space-time correlation contours can be approximated as ellipses when the EM1 parameters are used, thereby explaining the effectiveness of $r_E$ scaling in Fig. \ref{fig:10}. On the other hand, if $\tilde U_e$ and $\tilde V$ are estimated from the EM2 method and the range of $r_E$ values are kept the same as in Fig. \ref{fig:12}, one could see that the elliptic fits do not explain every feature of the $R_{TT}(r,\tau)$ contours (Fig. \ref{fig:13}). In particular, for all the five cable sections, the orientation of the parametric ellipses appear steeper than the original $R_{TT}(r,\tau)$ contours. This occurs because the EM2 method yields a larger $\tilde U_e$ value than EM1, eventually causing a poorer collapse of $R_{TT}(\tau,r)$ curves in Fig. \ref{fig:10}. 

Despite the success of EM1, one might ask why the collapse of curves in Fig. \ref{fig:10} with $r_E$ scaling was not a perfect one. From Figs. \ref{fig:12} and \ref{fig:6}e, one could infer that the  $R_{TT}(r,\tau)$ contours are approximately self-similar, meaning they all share a same preferential direction and aspect ratio. This self-similarity implies that the eddies of a particular scale are convected and swept past a fixed point analogously as the eddies of a larger scale. However, upon closely observing these contours, the two innermost ones appear distorted with respect to others. A plausible reason could be the range of spatial scales sampled in our flow encapsulate two scaling regimes displaying $-1$ ($\kappa z <$ 1) and $-5/3$ (1 $<\kappa z <$ 3) spectral slopes, respectively, thereby disrupting the perfect scale similarity. Note that for our dataset both of these scaling regimes do not span over an extended range of wavenumbers (Fig. \ref{fig:8}c) and therefore it remains difficult to ascertain this hypothesis. Moreover, for reasons that remain unknown, the fits for cable section 3 in Fig. \ref{fig:12} appeared somewhat poor, as the empirically determined $R_{TT}(r,\tau)$ contours exhibited slightly steeper tilts than the fitted ellipses. Owing to these anomalies, the collapse of curves in Fig. \ref{fig:10} with $r_E$ scaling was not a perfect one.



\subsubsection{Differences between EM1 and EM2}
The collapse of space-time correlation curves and the shapes of their iso-correlation contours suggest that the relationship between space and time in a heterogeneous forest clearcut flow can be described by an elliptic model when the turbulence intensities are large and sweeping effects are present. However, the efficacy of such description depends on what method has been used to compute the elliptic model parameters. For instance, since the convective speed estimates from the EM2 method are larger than EM1, difficulties arise while explaining the orientation of the space-time correlation contours. We hypothesize that these large $\tilde U_e$ values are related to a measurement limitation. 

For our set up we are limited in the spatial domain by the length of the cable at each section, which is nearly 100 m. This implies, spatially, the eddies larger than 100 m were not sampled. However, the temporal signatures of those eddies persisted in the time domain. This is evident from the spectral comparison presented in Fig. 8. The temporal spectra, as opposed to the spatial one, displayed an extended region of $-1$ spectral slope with a clear break at even larger scales (see Fig. S5 in the Supplementary Material for a demonstration). Such mismatch in information between the two domains creates an issue while applying the EM2 method. For any half-hour period and corresponding to any cable section, if one notices the $R_{TT}(r,\tau)$  and $R_{TT}(\tau,r)$ curves of the spatially-smoothed DTS data (Figs. S6b and d in the Supplementary Material), it is conspicuous that the $R_{TT}(r,0)$ values decrease more rapidly than the $R_{TT}(\tau,0)$ ones. As a result, the $\tau_0$ estimates of the EM2 method suffer from a bias since those are evaluated from the condition that $R_{TT}(0,r)= R_{TT}(\tau_0,0)$ holds in the temporal domain. It remains a question if this bias could be reduced, and to address that we conducted an experiment. We do not show the results from this experiment and only discuss their main features.

In this experiment, before evaluating the $R_{TT}(\tau,r)$ curves, we computed the temporal fluctuations by removing the spatial-averaged values of temperature from each instant of time instead of just subtracting a constant temporal mean. The temporal variations in spatial means were caused by eddies whose sizes were larger than the length of the cable. Therefore, by removing the time-dependent spatial means from the time series, one removes the temporal fingerprints of eddies larger than the cable length. As the mismatch between the temporal and spatial domain arises from these eddies, this experiment was designed to attenuate that discrepancy. Of course, due to random sweeping events, it is impossible to remove them completely so the results must only be interpreted in an approximate sense. 

Be that as it may; after carrying out this exercise, we calculated the parameters of the elliptic model by applying both EM1 and EM2 methods. Since we kept the spatial fluctuations as it is, the $U_e$ estimates were essentially similar, but $\tilde U_e$ behaved differently than $U_e$. In fact, rather than being larger, $\tilde U_e$ values now appeared smaller than $U_e$ while $\tilde V$ and $V$ estimates were nearly the same. However, since the large scale information was removed, the sweeping speed estimates decreased and did not show a 1:1 relationship with the turbulence kinetic energy. This outcome indirectly suggests that the random sweeping effects we observed in our flow could possibly be associated with eddies of scales larger than 100 m. 

More importantly, this experiment illustrates that whenever one has limited spatial information as compared to the temporal domain (which is often the case), differences in the convective speed estimations will be present between EM1 and EM2. Moreover, depending on how strong the mismatch is, this bias could either be on the larger or smaller side, and cannot be guessed beforehand. To check how important this bias is, it is recommended to compare the EM1 and EM2 estimates by investigating the collapse of the space-time curves and the elliptic fits obtained from them while describing the space-time iso-correlation contours. Notwithstanding this limitation, there is a silver lining. As discussed by \citet{hogg2013reynolds}, the EM2 method is useful in the case of a nearly non-existent mean wind flow, since in such conditions, the $r_p$-$\tau$ approach could give erroneous results due to very small $U_e$ values. In those situations, EM2 method provides a practical way to compute the sweeping speeds by using data from temporal domain, collected at a few specific spatial points. In that respect, it is encouraging to see that our results demonstrate the sweeping speed estimations are not very different (at least in a statistical sense) when compared against EM1 and EM2.

\subsection{Impact on EC measurements}
\begin{figure*}[h]
\centering
\includegraphics[width=1\textwidth]{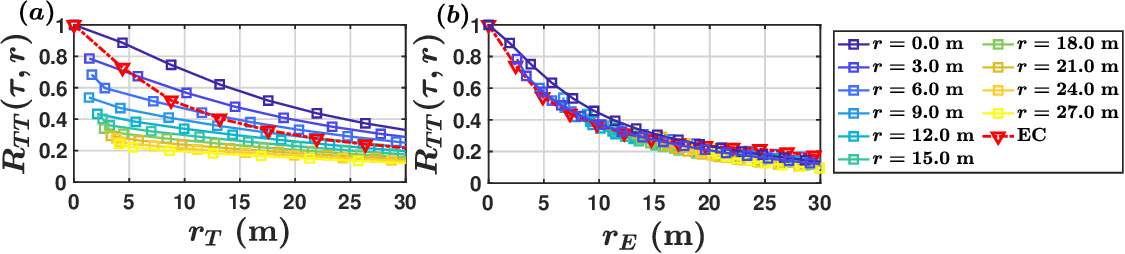}
\caption{Corresponding to cable section 4, the temporal cross-correlation curves ($R_{TT}(\tau,r)$) from the spatially smoothed DTS data are compared against the temporal auto-correlation curves of EC temperature measurements ($R_{TT}(\tau,0)$). The temporal lags are converted to spatial scales using (a) TH ($r_T$), and (b) EM1 ($r_E$) methods, respectively. Note that the 10-Hz EC measurements were block averaged to 2.3-s resolution in order to match with the DTS data.}
\label{fig:11}
\end{figure*}
Before drawing our final conclusions, we briefly comment on whether the random sweeping effects impact the EC measurements, located at the intersection of DTS cable section 4. In order to test this, we selected the EC temperature data corresponding to the periods for which the results are shown in Fig. \ref{fig:10}d (i.e. for section 4). Since the EC data are at a much finer temporal resolution (0.1 s), we block averaged them to 2.3 s resolution in order to match with the DTS data. After carrying out this block averaging, the auto-correlation curves of EC temperature were computed, and the temporal lags were converted to spatial scales using both Eqs. \ref{es} and \ref{TH}, i.e. by applying TH and EM. As the EM1 method produced the best collapse in Fig. \ref{fig:10}d, we only used the $(U_e,V)$ values from section 4 to execute the conversion for the EC data. These converted auto-correlation curves, when overlaid on the DTS derived $R_{TT}(\tau,r)$ curves, it is apparent that the $r_E$ scaling, as opposed to $r_T$, could bring them very close to the DTS ones (Fig. \ref{fig:11}). This result demonstrates that the EC temperature measurements are indeed impacted by the random sweeping events. Since DTS only measures temperature, we could not carry out a similar comparison for the velocity and other scalars.

\section{Conclusion}
\label{conclusion}
In this study, we have developed a comprehensive framework to test whether an elliptic model of space-time correlations is more appropriate than Taylor's hypothesis (TH) to describe the relationship between space and time in a highly heterogeneous forest clearcut flow. Our flow situation is symptomatic of a heterogeneous roughness sublayer flow, rather than a canonical atmospheric surface layer one. To execute our objective, an extensive dataset encompassing both Distributed Temperature Sensing (DTS) and Eddy Covariance (EC) measurements were used. These datasets were collected during a five-month field campaign (May-September 2024) and we restricted ourselves to buoyancy-dominated conditions when the mean wind speed was parallel to the forest edge. These wind scenarios were highly turbulent with turbulence intensities of the streamwise wind fluctuations exceeding 0.48. 

For this highly turbulent forest clearcut flow, the departure from TH was primarily caused by sweeping effects, which were found to be of the same order as the convective speeds of the temperature structures. These sweeping effects were associated with large scale eddies (possibly of scales larger than 100 m), and in accordance with the random sweeping hypothesis, their magnitudes scaled perfectly with the turbulence kinetic energy of the flow. In fact, their strong presence introduced curvatures in the shapes of the space-time correlation contours by making them appear as closed-form curves, rather than a family of straight lines as predicted by TH. To confirm whether these closed-form curves were indeed ellipses, the elliptic scaling was invoked. 

If the relationship between space and time could be expressed as ellipses, this particular scaling should collapse the temporal cross-correlation curves to a single curve when evaluated between points separated in space. This expectation was nearly fulfilled for our forest clearcut flow, as the elliptic scaling collapsed the space-time correlation curves of temperature fluctuations reasonably well over all the five DTS cable sections, positioned at distances one to six times of canopy height from the forest edge. The departure from a perfect collapse was related to the fact that the shapes of the space-time correlation contours at their largest values were distorted with respect to an ellipse. The exact reason behind this distortion is not clear. We hypothesize that it occurs because the range of spatial scales we sample in our flow encapsulate two scaling regimes with $-1$ and $-5/3$ spectral slopes, respectively, which, potentially, could disrupt the self-similarity of the space-time contours. 

We also compared our results against the two methods that exist to compute the parameters of the elliptic model. One method utilizes the linear relationships between the peak positions of temporal (spatial) cross-correlation curves and spatial (temporal) separations (EM1 method). The second method uses the information available from the temporal cross-correlation curves at different spatial separations to compute the same (EM2 method). For our case, the convective speed estimations from the EM2 method yielded larger values than EM1 while the sweeping speed estimations remained mostly similar. To evaluate the efficacy of the two, we applied the elliptic scaling on temporal cross-correlation curves by using the parameters estimated from both EM1 and EM2 methods. The results demonstrated that the EM1 method, not EM2, produced a better collapse of the temporal cross-correlation curves. 

This difference between the two methods is non-trivial and appears to be related to a mismatch between information available in spatial and temporal domain. For our experimental set up, we were limited in the spatial domain by the length of the cable at each section, which was nearly 100 m. This ensured that in the spatial domain the eddies of sizes larger than 100 m were not sampled. However, the temporal signature of those eddies persisted in the time domain. This disparity biased the convective speed estimations from the EM2 method. Therefore, one needs to be cautious while applying the EM2 method on a dataset that has limited spatial information, especially in high-Reynolds number flows (such as atmospheric boundary layer) encompassing a broad range of scales. Before trusting the EM1 and EM2 estimates, it is recommended to check if those could produce satisfactory collapse of the space-time curves and if the elliptic fits obtained from them could describe the morphology of the space-time contours fairly well. 

To summarize, the relationship between space and time in a buoyancy-dominated heterogeneous forest clearcut flow can be described as an ellipse when the turbulence intensities are large and random sweeping effects are present. We obtained a good agreement between the EC temperature auto correlation curves and DTS-derived temporal cross-correlation curves when the time to space conversion was achieved through the elliptic model. This agreement implies that the routine eddy covariance measurements are impacted by the random sweeping events in highly turbulent conditions. Nevertheless, it is not trivial to predict how these events would influence the flux measurements. The turbulent fluxes are a product of two signals and due to these sweeping events, their instantaneous values could randomly alternate between positive and negative depending on whether the constituent signals are in phase or out of phase. The resultant effect could be a decrease or increase in the time-averaged flux values, contingent upon how many of such random sweeping events have been sampled within a half-hour period. 

At present, these results are limited by a single observation height. One another drawback is, the DTS spatial temperature spectra at higher wavenumbers departed from the expected $-5/3$ slope, possibly due to noise, which we removed by a spatial smoothing operation. This, however, reduced the spatial resolution from 30 cm to 3 m. Accordingly, the estimations of the elliptic model parameters were based on an imperfect temperature signal with an additional constraint being imposed by the length of the cable itself. Due to this constraint, it remains unclear if there exists any upper bound on scales beyond which the elliptic model fails. Moreover, since DTS only measures temperature, one cannot know if the elliptic model holds for other scalars and velocity components. Therefore, for future studies, the following important questions need to be addressed:
\begin{enumerate}
    \item Do the structures in velocity and temperature move at similar speeds? If any dissimilarity exists between them, is that connected to the turbulent Prandtl number?
    \item Are sweeping effects and its scaling relationship with turbulence kinetic energy different between velocity and temperature?
    \item What is the relationship between the ratios of sweeping and convective speeds and turbulence intensities? Is there any threshold on turbulence intensity beyond which the sweeping effects cease to be important and TH becomes approximately valid?
    \item Is it possible that the elliptic model holds for one variable but not for the other? More importantly, is there any upper bound on scales up to which it holds?
    \item Can one extend the elliptic model to explain the spatial scales associated with turbulent fluxes of heat, momentum, and other scalars? 
\end{enumerate}

To answer them, the spatial data should span multiple heights, turbulence intensities, surface roughness (forest canopies, grasslands, arid deserts, water bodies, etc.), and signal types (velocity and scalars). For avoiding the sampling bias between spatial and temporal domain during convective conditions, it is advisable to have cable lengths as long as a kilometer. This definitely presents a challenge from a logistics perspective as well as finding a suitable location. Considering all these, it appears that the observations alone will not suffice, and large eddy simulations would be needed to complement the DTS measurements.

\acknowledgments
SC and OP thank the support from the Research Council of Finland (grant no. 354298). IM thanks support from ICOS-Finland via University of Helsinki funding. The authors thank Eduardo Martínez-García for providing the forest leaf area index data, Anssi Liikanen for helping with the field work and Pavel Alekseychik for the drone data used in Fig. \ref{fig:map}b.

%
%
\datastatement
The DTS and EC datasets are available at \citet{Chowdhuri2025dataset}. 

\authorcontribution
OP and SC designed the study. SC conceived the idea, carried out the analyses, developed the theoretical framework, prepared majority of the figures, and wrote majority of the manuscript. OP was responsible for funding acquisition, supervision and field measurements, conducted DTS data processing and provided the experimental data, wrote parts of the text, created Fig. \ref{fig:map} and edited the manuscript. IM provided minor edits to the manuscript.








%



\appendix[A] 
\label{app_A}
\appendixtitle{Assessment of stability}


\begin{figure*}[h]
\centering
\includegraphics[width=1\textwidth]{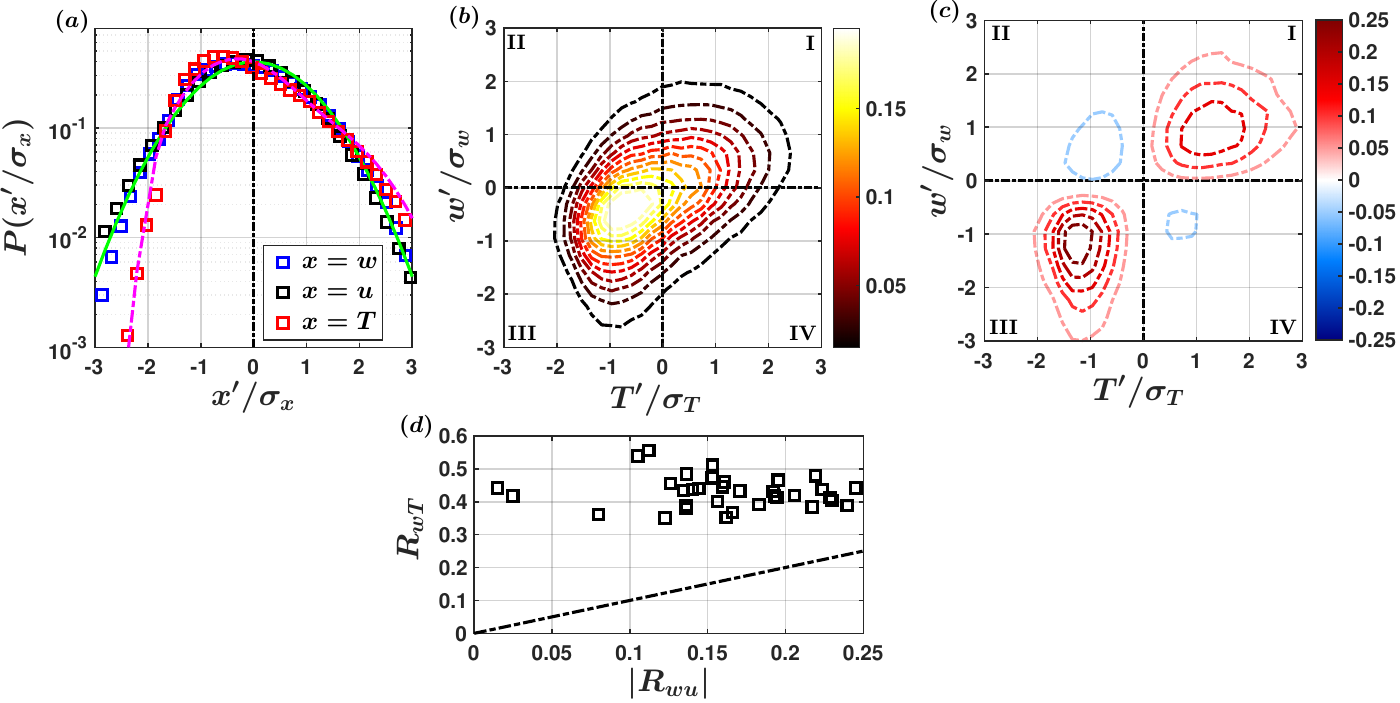}
  \caption{For the along-wind half hour periods from the EC clearcut measurements, (a) the PDFs of velocity and temperature fluctuations are shown. The green dash-dotted line indicates the Gaussian distribution, while the pink dash-dotted line represents the Gram-Charlier distribution. (b) The JPDFs between the vertical velocity and temperature fluctuations are shown where the colorbar represents the contour levels. (c) The flux-weighted JPDFs between the same two signals are shown. (d) The scatter plot between the correlation coefficients $R_{wT}$ and $R_{wu}$ are shown for along-wind cases. The absolute values of $R_{wu}$ ($|R_{wu}|$) are used and the dash-dotted line indicates the 1:1 relationship.}
\label{fig:A1}
\end{figure*}

In this appendix, we report the statistics of velocity and temperature fluctuations from the EC measurements corresponding to the along-wind conditions. These statistics are used to assess the stability conditions. In practice, the Obukhov similarity parameter is commonly used to determine the atmospheric stability, but we deliberately refrained from that approach since this theory was developed for homogeneous flows and its validity is questionable over heterogeneous surfaces such as a forest clearcut. In Fig. \ref{fig:A1}a, we show the PDFs of velocity and temperature fluctuations. Typical to a buoyancy-driven flow, the PDFs of temperature fluctuations remain strongly non-Gaussian as opposed to the velocity ones ($u^{\prime}$ and $w^{\prime}$). In fact, the PDFs of temperature fluctuations can be represented through a cumulant expansion method, otherwise known as Gram-Charlier distribution \citep{chowdhuri2019evaluation}. This distribution can be expressed as:
\begin{equation}
    P(T^{\prime}/\sigma_{T})=\frac{1}{\sqrt{2\pi}}\exp[-{(\frac{T^{\prime}}{\sigma_{T}})}^2][1+\frac{\mathcal{S}(T)}{6}(\mathcal{S}(T)-3\frac{T^{\prime}}{\sigma_{T}})],
    \label{gc}
\end{equation}
where $\mathcal{S}(T)$ is the skewness of temperature fluctuations. The non-Gaussianity in temperature statistics is also reflected in how the JPDFs between the vertical velocity and temperature fluctuations behave. From Fig. \ref{fig:A1}b, it is apparent that the JPDFs are strongly skewed towards quadrant III, which represents the downdraft motions. To further investigate the heat transport characteristics for our selected runs, we  computed the flux-weighted JPDFs, expressed as, $w^{\prime}T^{\prime}P(w^{\prime},T^{\prime})$, where $P(w^{\prime},T^{\prime})$ is the JPDF between $w^{\prime}$ and $T^{\prime}$. The contours of this quantity are plotted in Fig. \ref{fig:A1}c, and the area under them is proportional to the amount of heat being transported \citep{nakagawa1977prediction}. It is immediately clear from Fig. \ref{fig:A1}c that the transport of heat is strongly dominated by the down-gradient quadrants (I and III) with very small amounts being transported in counter-gradient directions (II and IV). 

This organization results in a situation where the heat is transported more efficiently than momentum. One could verify this argument in Fig. \ref{fig:A1}d, where the correlation coefficients between $w^{\prime}$ and $T^{\prime}$ ($R_{wT}$) remain considerably larger than the ones between $w^{\prime}$ and $u^{\prime}$ ($R_{wu}$). Since $R_{wu}$ values are negative, their absolutes ($|R_{wu}|$) are used on the $x$ axis of Fig. \ref{fig:A1}d. Additionally, the $R_{wT}$ values are nearly constant at around $0.48$, which is very close to $0.52$ and can be derived from free-convective scaling relations for the vertical velocity and temperature fluctuations \citep{wyngaard2010turbulence}. Therefore, by combining all these insights, one can claim that our selected periods were prototypical of a buoyancy-driven flow. 

\appendix[B] 
\label{app_B}
\appendixtitle{The role of large scale anisotropy}
\begin{figure*}[h]
\centering
\includegraphics[width=0.7\textwidth]{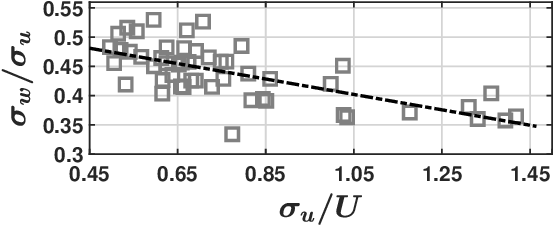}
  \caption{The scatter plot is shown between the streamwise turbulence intensities ($\sigma_u/U$) and the ratio $\sigma_w/\sigma_u$, a measure of large scale anisotropy. These quantities are estimated from half-hourly EC measurements corresponding to periods when the mean wind was parallel to the forest edge. The gray squares indicate individual data points while the black dash-dotted line shows the best-fit straight line between $\sigma_u/U$ and $\sigma_w/\sigma_u$.}
\label{fig:A2}
\end{figure*}

In Figs. \ref{fig:6}d and \ref{fig:10}b, we observed that the $V/U_e$ ratios were of the order of unity and they did not show a strong dependence on the streamwise turbulence intensities. This, at a first glance, may appear counter-intuitive as we find $U_e \propto U$ and $V \propto {(\rm TKE)}^{1/2}$ with the proportionality constants being 0.7 and 1, respectively. Since the TKE is mostly carried by large scale eddies and $\sigma_u$ is known to be affected by such eddies, one might expect that the $V/U_e$ would show a dependence on $\sigma_u/U$. However, this expectation is not fulfilled due to large scale anisotropy. To explain that, let us write $V/U_e$ as,
\begin{equation}
\frac{V}{U_e} \propto \frac{{(\rm TKE)}^{1/2}}{U},
    \label{b1}
\end{equation}
where we invoke the observations that $U_e \propto U$ and $V \propto {(\rm TKE)}^{1/2}$. Now, Eq. \ref{b1} could be rewritten as,
\begin{equation}
\frac{V}{U_e} \propto \frac{1}{U}\sqrt{\frac{\sigma_u^2+\sigma_v^2}{2}}\sqrt{1+\frac{\sigma_w^2}{\sigma_u^2+\sigma_v^2}},
    \label{b2}
\end{equation}
by using the fact that TKE=$(\sigma_u^2+\sigma_v^2+\sigma_w^2)/2$. By assuming $\sigma_u=\sigma_v$, Eq. \ref{b2} simplifies as,
\begin{equation}
\frac{V}{U_e} \propto \frac{\sigma_u}{U}\sqrt{1+\frac{\sigma_w^2}{2\sigma_u^2}}.
    \label{b3}
\end{equation}
The ratio $\sigma_w/\sigma_u$ is considered as a measure of large scale anisotropy. Since $\sigma_u$ is at the denominator, $\sigma_w/\sigma_u$ will decrease with increasing $\sigma_u$, thereby making the flow more anisotropic. This can be verified if one does a scatter plot between $\sigma_u/U$ and $\sigma_w/\sigma_u$. We show this scatter plot from EC measurements corresponding to periods when the wind was parallel to the forest edge (see Fig. \ref{fig:A2}). As expected, the anisotropy indeed increases with increasing streamwise turbulence intensities. Therefore, the term $\sqrt{1+\sigma_w^2/2\sigma_u^2}$ is inversely proportional to $\sigma_u/U$. Because of this inverse relationship, the product between the two remain largely constant and hence the $V/U_e$ ratios do not increase substantially with $\sigma_u/U$.

\bibliographystyle{ametsocV6}
\bibliography{references}

\end{document}